\documentclass[entropy,review,accept,pdftex,moreauthors]{Definitions/mdpi} 
\firstpage{1} 
\pubvolume{1}
\issuenum{1}
\articlenumber{0}
\pubyear{2025}
\copyrightyear{2025}
\externaleditor{Eduard Jorswieck} 
\datereceived{26 February 2025} 
\daterevised{3 April 2025} 
\dateaccepted{18 April 2025} 
\datepublished{ } 
\hreflink{https://doi.org/} 

\usepackage{comment}

\Title{A Survey on Semantic Communications in Internet of Vehicles}

\TitleCitation{A Survey on Semantic Communications in Internet of Vehicles}

\Author{\hl{Sha Ye $^{1}$,} 
 Qiong Wu $^{1,}$*\orcidA{}, Pingyi Fan $^{2}$\orcidB{} and Qiang Fan $^{3}$}

\AuthorNames{Sha Ye, Qiong Wu, Pingyi Fan and Qiang Fan}

        \AuthorCitation{\hl{Ye,} 
 S.; Wu, Q.; Fan, P.; Fan, Q.}

\address{%
$^{1}$ \quad School of Internet of Things Engineering, Jiangnan University, Wuxi 214122, China; shaye@stu.jiangnan.edu.cn\\
$^{2}$ \quad \hl{Department} of Electronic Engineering, State Key laboratory of Space Network and Communications, Beijing National Research Center for Information Science and Technology, Tsinghua University, Beijing 100084, China; fpy@tsinghua.edu.cn\\ 
$^{3}$ \quad Qualcomm, San Jose, CA 95110, USA; qf9898@gmail.com}

\corres{Correspondence: qiongwu@jiangnan.edu.cn; Tel.: +86-0510-8591-0633}

\abstract{The Internet of Vehicles (IoV), as the core of intelligent transportation system, enables comprehensive interconnection between vehicles and their surroundings through multiple communication modes, which is significant for autonomous driving and intelligent traffic management. However, with the emergence of new applications, traditional communication technologies face the problems of scarce spectrum resources and high latency. Semantic communication, which focuses on extracting, transmitting, and recovering some useful semantic information from messages, can reduce redundant data transmission, improve spectrum utilization, and provide innovative solutions to communication challenges in the IoV. This paper systematically reviews state-of-the-art semantic communications in the IoV, elaborates the technical background of the IoV and semantic communications, and deeply discusses key technologies of semantic communications in the IoV, including semantic information extraction, semantic communication architecture, resource allocation and management, and so on. Through specific case studies, it demonstrates that semantic communications can be effectively employed in the scenarios of traffic environment perception and understanding, intelligent driving decision support, IoV service optimization, and intelligent traffic management. Additionally, it analyzes the current challenges and future research directions. This survey reveals that semantic communications have broad application prospects in the IoV, but it is necessary to solve the real existing problems by combining advanced technologies to promote their wide application in the IoV and contributing to the development of intelligent transportation systems.}

\keyword{internet of vehicles; semantic communication; semantic information; semantic encoding and decoding; resource optimization} 

\usepackage{xcolor,soul} 
\renewcommand{\hl}[1]{#1} 
\begin{document}




\section{Introduction}

In today's digital era, the Internet of Vehicles (IoV) \cite{kaiwartya2016internet} and semantic \linebreak  communications \cite{luo2022semantic,gunduz2022beyond,shi2021semantic}, as cutting-edge technologies in the field of intelligent transportation and communication, are gradually changing the way we travel and are transforming traffic management. The IoV realizes real-time sharing and interaction of traffic information through vehicle-to-vehicle (V2V), vehicle-to-infrastructure (V2I), and vehicle-to-pedestrian (V2P) communications. It greatly improves traffic efficiency and safety. Semantic communications, on the other hand, are an emerging communication paradigm that significantly improves the efficiency and reliability of communications by directly transmitting the semantic content of information instead of the traditional sequence of bits.

However, with the continuous emergence of new applications in the IoV, such as autonomous driving \cite{duan2020emerging}, vehicle remote monitoring and diagnostics~\cite{maksimychev2021connected}, and IoV entertainment services \cite{ang2018deployment}, the mobile data traffic between vehicles and roadside units (RSUs), as well as between vehicles, has shown an explosive growth trend. This surge in data volume has extremely stringent demands on the communication infrastructure of the IoV, particularly highlighting the issue of scarce spectrum resources. When facing massive amounts of data, traditional IoV communication technologies often reveal limitations such as limited spectrum resources, high latency, and insufficient capabilities for processing semantic information. These issues significantly restrict the further enhancement of IoV performance and hinder the development of intelligent transportation systems.

Against this background, semantic communication becomes a promising  communication technology to address the IoV communication bottleneck. Unlike traditional communication that  primarily focuses on the accurate transmission of data, semantic communication concentrates on extracting, transmitting, and recovering the semantic information contained within messages, enabling the receiver to understand the intent and content of the sender. \hl{The} 
 integration of the IoV and semantic communication can effectively improve the efficiency of data transmission and ensure the accuracy and reliability of information transfer in complex traffic environments. The concealment of semantic communication provides a new approach for privacy protection in the Internet of Vehicles, making information transmission more secure. At the same time, it can flexibly adjust information content and services according to the specific needs and scenarios of users, providing more precise support for applications such as autonomous driving and intelligent traffic management.

{This combination also faces some challenges. For example, in dynamic and changeable traffic scenarios, how to accurately understand and process contextual information and how to ensure the efficient decoding and comprehension of information while maintaining its concealment are all issues that require further research.
Despite these challenges, the combination of the IoV and semantic communication undoubtedly paves a new way for the development of future intelligent transportation, and its potential and value are worthy of in-depth exploration. }

\begin{figure}[H]
	\includegraphics[width=13.5 cm]{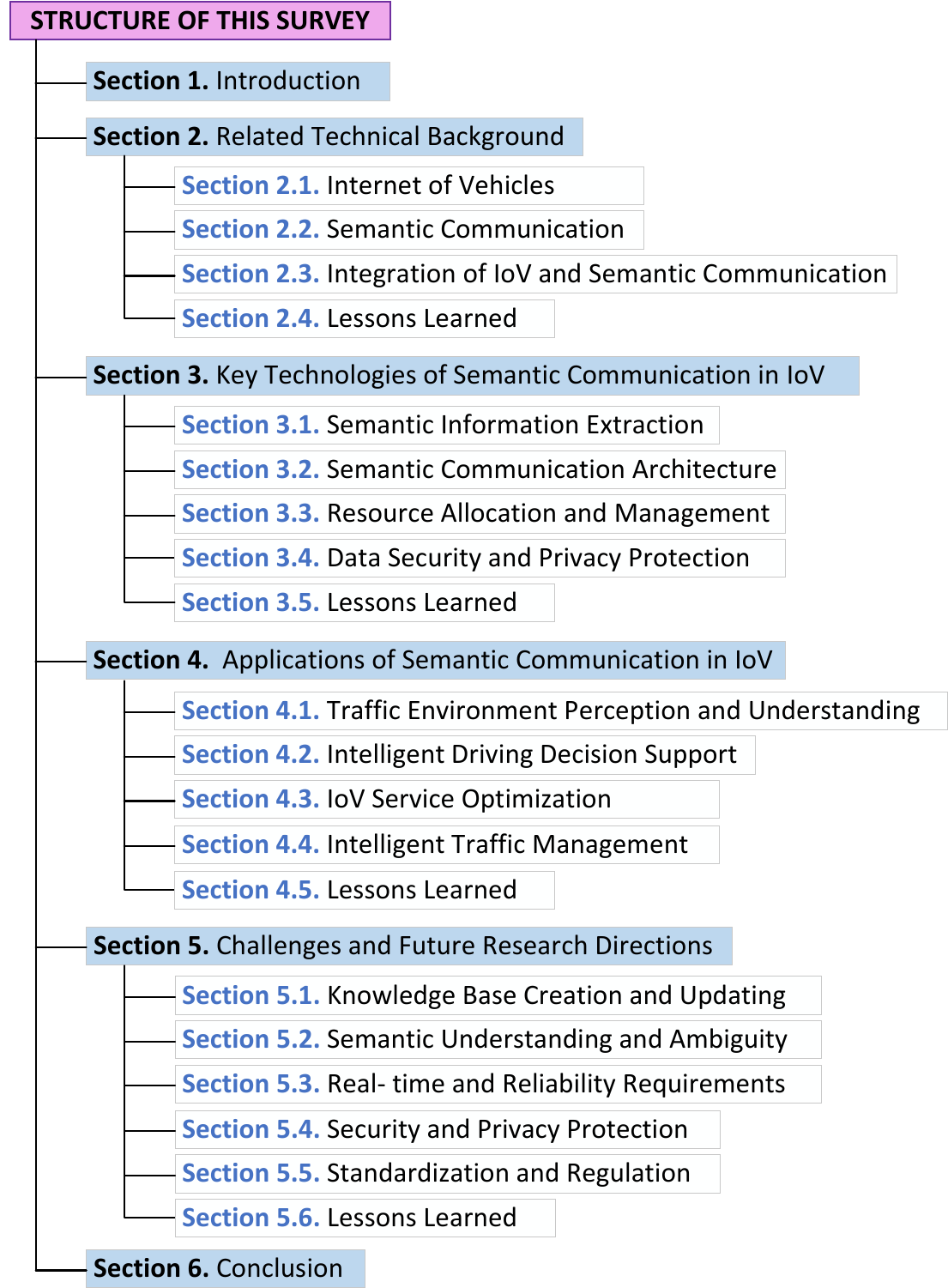}\vspace{5pt}
	\caption{\hl{Structure} 
 of this survey.\label{fig1}}
\end{figure}

In recent years, researchers have paid attention to the application of semantic communication in the field of the IoV. A substantial amount of research has been conducted on how to integrate the advantages of semantic communication into the IoV. In light of this, this survey provides a comprehensive and systematic review of the current research related to semantic communications in the IoV. We summarize these studies from two perspectives: key technologies and specific applications, as shown in Table ~\ref{tab1}. To the best of our knowledge, this is the first comprehensive study discussing the IoV based on semantic communication.The rest of the paper is organized as follows: Section \ref{sec2} begins with an overview of the architecture and key technologies of the IoV, as well as the fundamental principles, system architecture, and classification of semantic communication, providing the theoretical background for subsequent research.  Section \ref{sec3} delves into the key technologies of semantic communication in the IoV, covering the core technical areas of semantic information extraction, semantic communication architecture, resource allocation and management, and details regarding current research progress. Section \ref{sec4} demonstrates the significant achievements of semantic communication in practical application scenarios such as traffic environment perception and understanding, intelligent driving decision support, IoV service optimization, and intelligent traffic management through specific case studies, further validating the potential of semantic communication in the IoV. Section \ref{sec5} provides an in-depth analysis of the current challenges and a prospective outlook on future research directions. Finally, in Section \ref{sec6}, we present the conclusions of this survey. Additionally, the organization and structure of the survey are shown in Figure~\ref{fig1}.

\begin{table}[H]
	\caption{Studies related to semantic communication in IoV.\label{tab1}}
		\begin{adjustwidth}{-\extralength}{0cm}
		\newcolumntype{C}{>{\centering\arraybackslash}X}
				\begin{tabularx}{\fulllength}{cCcC}
				\toprule
				\textbf{ }	& \textbf{Category}	& \textbf{Literatures }     & \textbf{Core Content}\\
				\midrule
				\multirow[m]{8.5}{*}{\centering Key Technologies}	& Semantic Information Extraction			& \cite{audebert2017segment,10.1007/978-981-16-0081-4_63,9304779,8551565,10118717,10574825,feng2024semantic,lu2024generative}			& Extract semantic information from multimodal data.\\
				\cmidrule{2-4}
				& Semantic Communication Architecture			& \cite{10118717,feng2024semantic,lu2024generative,10013090,eldeeb2024multi,hu2024semharq,lv2024importance,10588546,10473044,10217468,feng2023scalable,10506539,raha2023generative,10815060,LIN2024,10198474,10847825}			& Diverse semantic communication architectures covering multitask and multiuser collaboration, image transmission, generative AI, etc.\\
				\cmidrule{2-4}
				& Resource Allocation and Management			& \cite{10506539,10636300,10644092,shao2024semantic3,9538906,10073623,10681786,10570867,10416926,10333738}			& Resource allocation methods based on reinforcement learning, optimization theory, and federated learning.\\
				\midrule
				\multirow[m]{8.5}{*}{\centering Specific Applications}    & Traffic Environment Perception and Understanding			& \cite{imran2024environment,sheng2024semantic}			& Enhance the coverage and accuracy of environmental perception.\\
				\cmidrule{2-4}
				& Intelligent Driving Decision Support			& \cite{10622833,10049005}			& Extract key semantic information to support real-time decision making.\\
				\cmidrule{2-4}
				& IoV Service Optimization			& \cite{10271127,liu2024receivercentricgenerativesemanticcommunications,9788561}			& Optimize service performance through semantic communication.\\
				\cmidrule{2-4}
				& Intelligent Traffic Management			& \cite{10571211,10095672,zeng2024dmce,10719121}			& Improve traffic management efficiency and emergency response~capabilities.\\
				\bottomrule
			\end{tabularx}
			\end{adjustwidth}
	\end{table}

\section{Related Technical Background}\label{sec2}

\subsection{Internet of Vehicles }\label{sec2.1}

{The Internet of Vehicles (IoV) is an intelligent transportation system that connects vehicles with other vehicles, infrastructure, pedestrians, and roadside units via advanced communication technologies.} Essentially, it is a deep integration of vehicular ad hoc networks (VANETs) and the Internet of Things (IoT) \cite{8674992,sharma2019survey}. {It aims to enhance traffic safety, improve transportation efficiency, and optimize the driving experience through real-time information sharing and collaborative interaction~\cite{w12,w9}. It can be primarily categorized as~follows:}

\begin{itemize}
	\item\textcolor{black}{Vehicle-to-Vehicle (V2V)~\cite{ALSULTAN2014380}: Vehicles directly exchange information with each other, such as speed, direction, and emergency braking signals. This mode enables vehicles to anticipate potential collision risks in advance, thereby enhancing driving safety.}
	\item	\textcolor{black}{Vehicle-to-Infrastructure (V2I)~\cite{ALSULTAN2014380}: Vehicles communicate with roadside units, such as traffic lights and road sensors, to obtain information on traffic conditions and traffic signal status. V2I communication helps optimize traffic flow and reduce congestion.}
	\item	\textcolor{black}{Vehicle-to-Pedestrian (V2P)~\cite{tahmasbi2017implementation}: Vehicles interact with pedestrian devices, such as smartphones and smartwatches, to alert pedestrians to the approach of vehicles or to provide vehicles with pedestrian location information.}
	\item	\textcolor{black}{Vehicle-to-Roadside Unit (V2R)~\cite{wu2018computational}: Vehicles communicate with roadside units (RSUs), which are devices installed along roadsides to provide local traffic information, weather conditions, and other relevant data.}
	\item	\textcolor{black}{Vehicle-to-Device (V2D)~\cite{jomaa2017comparative}: Vehicles communicate with various mobile devices to offer personalized services, such as vehicle status monitoring and remote control.}
	\item	\textcolor{black}{Vehicle-to-Grid (V2G)~\cite{endo2018evaluation}: Vehicles interact with the power grid, functioning as mobile energy storage units that can exchange energy with the grid. For example, vehicles can charge during periods of low grid load and discharge during periods of high grid load, thereby improving the stability and efficiency of the power grid.}
\end{itemize}

\textcolor{black}{Through these communications, the IoV can achieve a smarter, safer, and more efficient transportation system. A mature IoV system typically incorporates all of the above communication modes.}


\subsubsection{IoV Architectures}
To address the diverse requirements of IoV scenarios, researchers have proposed various IoV architectural frameworks. The most common classifications include the three-tier architecture, the in-vehicle and intervehicle network architecture, and the vehicle--road--cloud integrated architecture.

\begin{figure}[H]
	\includegraphics[width=13 cm]{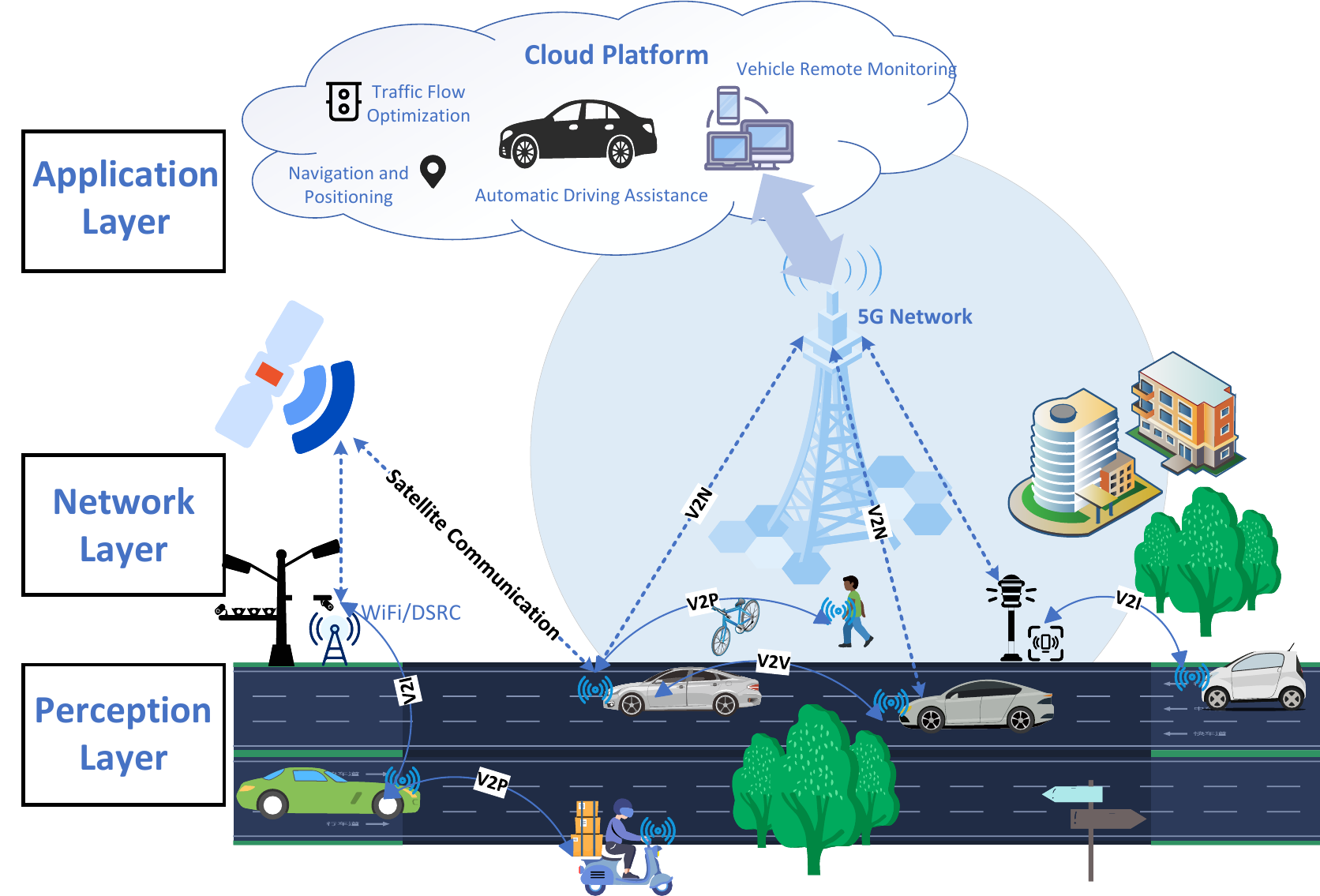}
	\caption{IoV scenario architecture diagram.\label{fig2}}
\end{figure}

The typical three-tier IoV architecture consists of the perception layer, network layer, and application layer~\cite{sadiku2018internet,9852810,panigrahy2023survey}, as shown in Figure~\ref{fig2}. The perception layer acts as the nerve endings of the IoV, utilizing technologies such as sensors, RFID, and GPS to collect information on vehicle status, road environment parameters, and surrounding vehicle dynamics, providing foundational data for the IoV, such as vehicle speed and tire pressure. The network layer serves as the hub for information exchange, leveraging cloud computing, virtualization, and other technologies to integrate various communication networks, such as 4G/5G, Wi-Fi, Bluetooth, and DSRC, enabling information sharing and interconnectivity between vehicles, roads, people, and the cloud. The application layer is the core value realization for the IoV, developing and optimizing functions such as intelligent traffic management, vehicle safety control, and traffic event warning based on cloud computing platforms, offering users services like vehicle information queries, information subscriptions, and event alerts. In addition to the three-tier architecture, researchers also proposed more detailed four-tier~\cite{8932226,sharma2019survey,8767077}, five-tier~\cite{kaiwartya2016internet}, and seven-tier~\cite{contreras2017seven,7892008} architectures for IoV scenarios. 

In terms of subnetwork division, IoV can be categorized into two main subnetworks: the in-vehicle network and intervehicle network~\cite{6969789,ALNASSER201952,TASLIMASA2023100809,GUO2024237}. The in-vehicle network utilizes bus technologies such as the Controller Area Network (CAN), the Local Interconnect Network (LIN), FlexRay, MOST, and Ethernet to construct the internal network of the vehicle, connecting sensors and Electronic Control Units (ECUs) within the vehicle to enable functions like vehicle status monitoring, fault diagnosis, and intelligent control. The intervehicle network consists of four entities: the onboard unit, roadside users, roadside units (RSUs), and cloud servers. It leverages technologies such as GPS, wireless local area networks, and cellular networks to build an open, dynamic mobile communication network, facilitating real-time information exchange between vehicles, vehicles and road infrastructure, and vehicles and pedestrians, thereby enhancing traffic safety and efficiency.

Furthermore, the IoV can also be divided according to the vehicle--road--cloud integrated architecture~\cite{6871677,9088328}, which primarily consists of three components: the onboard unit, road facility unit, and cloud control platform. Specifically, the onboard unit integrates sensors, control units, and communication devices within the vehicle to perceive vehicle status and environmental information and transmit it to the cloud. The road facility unit includes traffic lights, roadside cameras, and sensors, collecting traffic and environmental information from the road and communicating with the onboard unit to enable data exchange and sharing. The cloud control platform, composed of multiple cloud systems, undertakes tasks such as data storage, processing, and analysis, providing vehicles with real-time traffic information, road condition data, and navigation guidance. Simultaneously, it receives data uploaded by vehicles while offering a wide range of applications and services and maintaining end-to-end Quality of Service (QoS) for users~\cite{w1,w2}.

\subsubsection{Communication Technologies}
Communication technologies in the IoV include V2X communication technology, 4G/5G mobile communication technology, Dedicated Short-Range Communication (DSRC)~\cite{8967260}, Bluetooth, Wi-Fi, and others, enabling efficient information interaction between vehicles and the external environment. In \cite{sharma2019survey}, three types of wireless communication technologies are summarized: vehicular communication, cellular mobile communication, and short-range static communication: 
\begin{itemize}
	\item	Vehicular communication: In VANETs, On-Board Units (OBUs) and roadside units (RSUs) often use DSRC for communication. For example, vehicles transmit collected data such as speed, acceleration, and fuel levels to nearby RSUs or other vehicles via OBUs based on the IEEE 802.11p standard. RSUs also use DSRC to communicate with OBUs, enabling functions such as information forwarding, local communication, and road safety information provision. Continuous Air-interface for Long and Medium range (CALM) is another important vehicular communication standard in IoV, providing specifications and support for communication, ensuring compatibility and stability between vehicles and other devices.
	\item	Cellular mobile communication: Vehicles can communicate with other networks through 4G/LTE, WiMax, and other cellular network technologies, enabling V2I connectivity, thereby extending communication range and accessing more comprehensive traffic and environmental information~\cite{w3,w46}. Among these, 5G NR V2X, as a new-generation cellular IoV technology, offers ultra-low latency, high reliability, and high data transmission rates, significantly enhancing the performance and safety of IoV applications. It supports advanced V2X scenarios such as platooning, remote driving, and sensor extension. Satellite communication also plays a role in special scenarios or remote areas, ensuring vehicle connectivity. For instance, in mountainous regions or areas with poor signal coverage, satellite communication can maintain the vehicle's connection with the outside world.
	\item	Short-range static communication: This includes technologies such as Zigbee, Bluetooth, and Wi-Fi. Zigbee can be used for short-range communication between vehicles and sensors, enabling environmental perception and data collection, such as vehicle status monitoring and environmental parameter detection. Bluetooth is primarily used for connecting vehicles with personal devices (e.g., smartphones, tablets, etc.), facilitating device interaction and information sharing within the vehicle. Wi-Fi provides high-speed network access for vehicles in specific areas (e.g., parking lots, service stations, etc.), meeting the demand for large data transmission and real-time information acquisition. For example, vehicles in parking lots can download maps or access service information via Wi-Fi.
\end{itemize}

\subsection{Semantic Communication}
\textcolor{black}{Semantic communication is a method of encoding, transmitting, and decoding information based on its meaning and intent, aiming to achieve efficient communication by extracting and conveying the essential semantic content of the information. For instance, in an autonomous driving scenario where a vehicle needs to quickly identify and react to obstacles ahead, traditional communication methods might require the transmission of large amounts of image data. In contrast, semantic communication would simply transmit the critical message that “there is an obstacle ahead,” thereby saving time and bandwidth.}

\subsubsection{Semantic Communication System Architectures}
Weaver~\cite{shannon1948mathematical} proposed three levels of communication, including technical issues, semantic issues, and effectiveness issues. The technical layer focuses on the precise transmission of information between the sender and receiver; the semantic layer focuses on the meaning that the receiver understands and the meaning that the sender wants to express; and the effectiveness layer focuses on the effectiveness of the information to achieve the expected impact on the receiver's behavior. Traditional communication mainly focuses on the technical layer, pursuing the accurate transmission of data bits~\cite{w43,w47,7488201}. However, with the explosive growth of data volume in the era of big data, the transmission of a large amount of raw data by traditional communication methods leads to problems such as inefficient transmission and waste of resources~\cite{5720555}. Semantic communication, on the other hand, focuses on the latter two layers and aims to accurately transmit semantic meanings rather than mere bit information~\cite{10319671}. Semantic communication systems include components such as a semantic encoder, channel encoder, channel decoder, semantic decoder, and knowledge base~\cite{xin2024semantic}, \textcolor{black}{as shown in Figure~\ref{fig3}}.\vspace{-5pt} These components are described in the following: 

\begin{figure}[H]
	\includegraphics[width=14 cm]{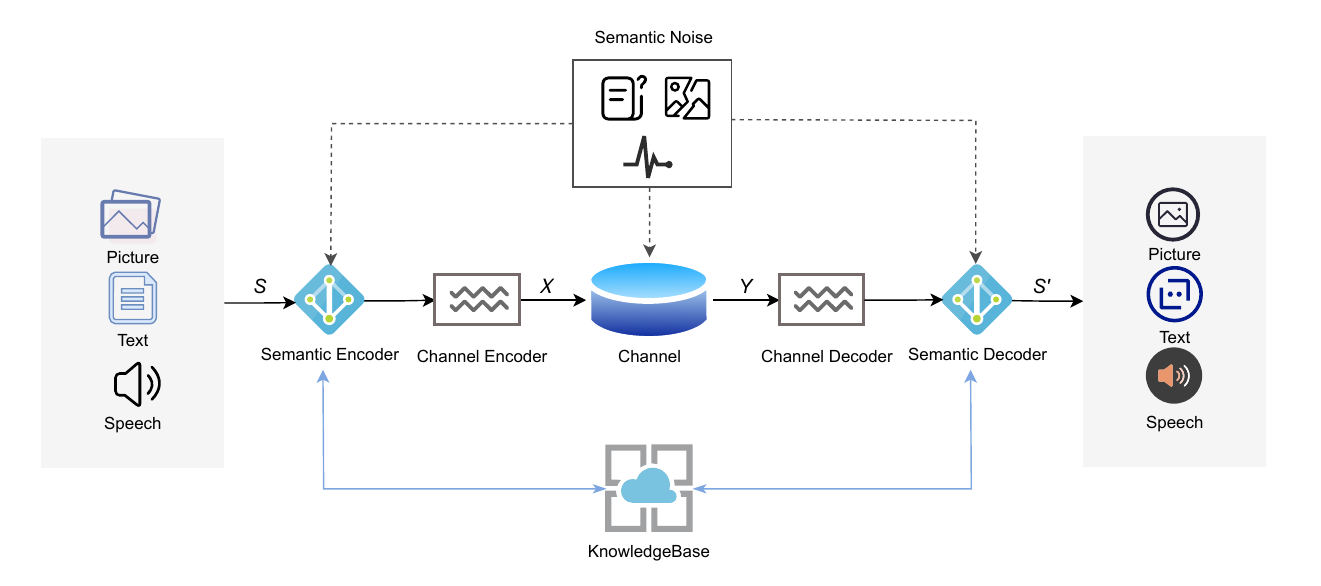}
	\caption{{Semantic communication system architectures.}\label{fig3}}
\end{figure}   

\begin{itemize}
	\item	Semantic encoder: The semantic encoder accurately extracts semantic information from the source message and effectively compresses it through a series of complex algorithms and techniques.
	\item	Channel encoder: The channel encoder encodes and modulates the semantic features processed by the semantic encoder and maps the semantic features into a signal form suitable for transmission over the channel.
	\item	Channel decoder: The channel decoder uses a decoding algorithm corresponding to the channel encoder to restore the encoded semantic features from the received signal.
	\item	Semantic decoder: The semantic decoder converts the signal output from the channel decoder into a format that the user can understand.
	\item	Knowledge base: The knowledge base provides the semantic understanding basis for the encoder and decoder. It further analyzes and transforms the recovered semantic features based on pre-established semantic rules and the knowledge system within the knowledge base. Depending on the scope of knowledge sharing and the user, it can be divided into local knowledge bases and public knowledge bases.
	\item	Semantic noise: Semantic noise is an interfering factor that leads to wrong recognition and interpretation of semantic information in the whole process of semantic communication. Semantic noise includes semantic and physical channel noise, such as text semantic ambiguity, image antagonistic samples, and so on.
\end{itemize}

\textcolor{black}{In a semantic communication system,} assume that the input raw data are \hl{S,} 
 which can be text, image, etc. After passing through the semantic encoder and the channel encoder, we can obtain the encoded data as 
\begin{linenomath}
	\begin{equation}
X = ch_{\beta}(se_{\alpha}(S)) , 
\end{equation}
\end{linenomath}
where \(se_{\alpha}(\cdot)\) and \(ch_{\beta}(\cdot)\) are the semantic and channel encoder networks with parameters \(\alpha\) and \(\beta\), respectively.

\textcolor{black}{After passing through the channel, }we can obtain
\begin{linenomath}
	\begin{equation}
Y = HX + N ,
\end{equation}
\end{linenomath}
 where H and N represent the channel gain and noise, respectively. The specific formula expression usually needs to be calculated according to the specific scenario.

\textcolor{black}{Then, after passing through the channel decoder and the semantic decoder, }we can reconstruct the data as
\begin{linenomath}
	\begin{equation}
S'=se_{\mu}^{- 1}(ch_{v}^{-1}(Y)) , 
\end{equation}
\end{linenomath}
where \(se_{\mu}^{-1}(\cdot)\) and \(ch_{v}^{-1}(\cdot)\) are the networks of the semantic and channel decoders with parameters \(\mu\) and \(v\), respectively.

Unlike the traditional communication paradigm of ``transmit first, then understand'', semantic communication adopts a ``understand first, then transmit'' paradigm, ensuring that it can overcome the limitations of traditional communication. Semantic information extraction is the foundation of semantic communication, which can extract semantic information from raw data and reduce the amount of transmitted data. This requires the precise extraction of key semantic elements from various complex data sources, such as text, speech, images, and sensor data. This relies on advanced technical methods, such as the artificial intelligence method~\cite{zhang2024advances}. For example, in the semantic extraction of textual information, Natural Language Processing (NLP) techniques can determine the core semantics of text through lexical and syntactic analysis, as well as semantic role labeling. For image data, computer vision techniques such as object detection, image classification, and semantic segmentation algorithms can identify objects and their relationships within images, thereby extracting corresponding semantic information. As for speech signals, they are first converted into text using speech recognition technology, and then their semantic connotations are further explored.

\clearpage 
In the design of semantic communication systems, the encoding process is crucial for achieving efficient semantic transmission. Similar to traditional communication systems, semantic communication systems also involve source coding and channel coding, but their design goals and methods differ significantly from those of traditional communication. In semantic communication, the source coding is primarily responsible for extracting semantic information from raw data, removing redundancy, and improving transmission efficiency. Channel coding, on the other hand, focuses on addressing noise and interference in the channel to ensure the reliability of semantic information during transmission. Traditional communication systems typically employ separate source coding and channel coding approaches. Source coding is responsible for redundancy removal, while channel coding enhances anti-interference capabilities. However, this separated design may not fully leverage the characteristics of semantic information in semantic communication. Therefore, semantic communication systems tend to adopt Joint Source--Channel Coding (JSCC), which jointly optimizes source and channel coding to meet the transmission requirements of semantic information.  

In recent years, the JSCC technology based on deep learning has shown significant advantages in semantic communication. For example, Deep JSCC technology~\cite{8723589} employs Convolutional Neural Networks (CNNs) to achieve end-to-end semantic information encoding and decoding for wireless image transmission, exhibiting excellent performance under low signal-to-noise ratios (SNRs) and limited bandwidth conditions. Further optimized Deep JSCC-f~\cite{9066966} introduces a channel output feedback mechanism and adopts a hierarchical autoencoder structure, showing progressive improvement and robust performance in low-SNR scenarios. In the field of text transmission, DeepSC~\cite{9398576} extracts semantic information from text based on the Transformer model and achieves efficient transmission through joint semantic--channel coding, outperforming traditional methods under low-SNR conditions. For speech signal transmission, DeepSC-S~\cite{9450827} combines attention mechanisms and Squeeze-and-Excitation (SE) networks to effectively extract semantic information from speech, significantly improving the relevant indicators in telephony and multimedia systems. Additionally, the MU-DeepSC system~\cite{9653664}, designed for multiuser and multimodal data transmission, jointly designs the transmitter and receiver using Deep Neural Networks (DNNs), focusing on transmitting and recovering task-relevant semantic information and providing a novel solution for semantic communication in complex scenarios.

\subsubsection{Semantic Communication Types}
Based on the roles of semantics in the communication process, communication objectives, and application scenarios, semantic communication can be categorized into three types: semantics-oriented communication, goal-oriented communication, and semantics-aware communication~\cite{9955312}.

Semantics-oriented communication: In this communication design, the key lies in the accuracy of the semantic content of the source data. Unlike traditional content-blind source coding, semantic communication introduces a semantic representation module before encoding, which is responsible for extracting the core information and filtering redundancy, and semantic encoding and semantic decoding are similar to source coding and source decoding, respectively, in traditional communication. In practical scenarios, such as image transmission, irrelevant image details are filtered before transmission according to different tasks to reduce the network burden without affecting system performance. At the same time, the communication parties need to share local knowledge in real time; otherwise, semantic noise will be generated, resulting in semantic ambiguity, even if the physical transmission is free of syntactic errors.

Goal-oriented communication: This type needs to capture the semantic information, where the communication goal plays an important role in the semantic extraction process and thus helps to further filter the irrelevant semantic information in each transmission. For example, image transmission tasks are different and require different image features to be transmitted. In contrast to semantics-oriented communication, the outputs are direct action execution instructions aimed at accomplishing the task with limited network resources rather than focusing only on semantic accuracy. Similarly, the local knowledge of the communicating parties and the goal of the communication need to be consistent, otherwise semantic noise may lead to task failure.

Semantic-aware communication: Semantic-aware communication plays a role in task-oriented communication in scenarios such as autonomous driving and drone swarms. In these scenarios, multiple intelligent agents accomplish tasks through centralized or distributed collaboration, and semantic-aware communication establishes multiple explicit or implicit connections between different endpoints to enhance knowledge sharing among agents. Its semantic information is obtained by analyzing the behavior of intelligent agents and current environment in which the task is performed rather than extracting from data sources. Unlike the above two types, there is no generalized system model, and it is not part of traditional connection-oriented communication.

\subsubsection{Semantic Metrics}
Semantic metrics mainly depended on the similarity, which can be divided into two aspects: the semantic level and effectiveness level. The semantic level focuses on the accurate understanding and transmission of information, while the effectiveness emphasizes the performance and functional achievement of the system in practical applications~~\cite{10319671}.  

Semantic-level metrics: \textcolor{black}{The measurement of semantic similarity often serves to evaluate the similarity between the previously mentioned original input data S and the reconstructed output data S'.} For different types of media such as text, images, and speech, there are various semantic similarity measurement methods. For example, text has average semantic distortion~\cite{6815245} and Bilingual Evaluation Understudy (BLEU)~\cite{papineni2002bleu}; for images, metrics such as the peak signal-to-noise ratio (PSNR)~\cite{8723589} and Structural Similarity Index (SSIM)~\cite{sara2019image} are used; for speech, metrics like the Signal-to-Distortion Ratio (SDR)~\cite{vincent2006performance} and Perceptual Evaluation of Speech Quality (PESQ)~\cite{941023} are employed,  where each one has it pros and cons. Additionally, there are metrics for evaluating data compression ratios, such as the bit reduction rate~\cite{9832831} and average bits~\cite{9791409}.  

Effectiveness-level metrics: These include metrics based on Age of Information (AoI)~\cite{9380899} and Value of Information (VoI) ~\cite{molin2019scheduling}. The core of AoI-based metrics lies in measuring the time interval from the generation of information to its reception, while VoI-related metrics focus on the contribution of information to the completion of specific tasks. These metrics are used to assess the contribution of information to task completion, with different metrics being suitable for different scenarios and tasks.

\subsection{Integration of IoV and Semantic Communication}
The IoV imposes stringent requirements on communication technologies in terms of reliability, low latency, high bandwidth, and security~\cite{w6,w7,w32}. For instance, in autonomous driving scenarios, vehicles need to acquire real-time and accurate traffic information to make safe decisions, which demands extremely low latency and high reliability in communication. In applications such as remote vehicle monitoring and diagnostics, the large volume of data transmission necessitates a high bandwidth. With the increasing number of vehicle smart sensors, growing IoV communication demands, and diversification of user needs, the data volume transmitted and processed in IoV has surged dramatically~\cite{w33}. Traditional communication technologies, such as cellular networks and DSRC, face limitations in meeting IoV communication requirements, including limited spectrum resources, increased latency during network congestion, and insufficient capability to process semantic information. Therefore, how to improve
 spectrum utilization efficiency, real-time performance, and accuracy in IoV communication, while ensuring communication security and privacy protection, are urgent challenges to address.  

Semantic communication offers several significant advantages~\cite{Sun2023}. First, it reduces the transmission of redundant data, improving spectrum utilization and enabling efficient information exchange. Second, it enhances communication effectiveness and accuracy through contextual information, particularly in environments with limited bandwidth or low signal-to-noise ratios. Third, semantic communication can obscure information, ensuring that only individuals with specific background knowledge can understand the transmitted content. Finally, it focuses on the meaning of information and user needs, allowing flexible adaptation to diverse communication scenarios. These characteristics provide vast potential for IoV applications, optimizing data transmission, enhancing driving safety, protecting user privacy, and supporting intelligent decision making, thereby advancing the development of intelligent transportation systems.  

\textcolor{black}{In recent years, the application of semantic communication technology in the IoV has received extensive attention. By extracting, transmitting, and restoring the semantic content in information, semantic communication significantly improves the efficiency and reliability of communication. The Telecommunication Standardization Sector of the International Telecommunication Union (ITU-T) released a technical report ~\cite{itu-t-2023} in November 2023, which thoroughly explored the application potential of the Semantic-Aware Network (SAN) in future networks, including the Internet of Vehicles. The report states that by adopting semantic terms and grammars shared by machines and humans, the SAN can effectively represent, annotate, analyze, and interpret network and user-generated data, thus supporting automatic data processing, learning, and analysis.  
From a vehicular network perspective, the SAN enhances interoperability by translating heterogeneous data (e.g., LiDAR, camera feeds, etc.) into unified semantic domains, supports real-time decision making through semantic cognition, and ensures reliability via distributed semantic resource~provisioning.}

The integration of the IoV and semantic communication has extensive application scenarios, including autonomous driving, intelligent traffic management, remote vehicle diagnostics and maintenance, and connected entertainment. Autonomous driving refers to the technology that enables vehicles to perform various driving tasks without continuous human intervention. The IoV provides the necessary technical support for autonomous driving by enabling real-time communication between vehicles and between vehicles and infrastructure. This allows autonomous vehicles to acquire information about their surroundings, make more accurate decisions, and improve driving safety and comfort. Semantic communication in the IoV can further enhance the efficiency and accuracy of information transmission, enabling vehicles to better understand and respond to complex traffic scenarios.  In intelligent traffic management, IoV technology connects vehicles with traffic infrastructure to enable smart traffic management, such as traffic signal control, road condition monitoring, and electronic toll collection, thereby improving traffic efficiency and alleviating urban congestion. Semantic communication can provide personalized traffic information and services based on vehicle needs and user preferences. For example, it can offer customized navigation suggestions, traffic updates, and information about nearby facilities based on the vehicle's route and destination, thus supporting the efficient operation and optimized management of intelligent transportation systems. Additionally, a semantic IoV can be applied to remote vehicle diagnostics and maintenance, as well as connected entertainments. In remote diagnostics and maintenance, vehicles can transmit data in real time to service centers to accurately identify faults, provide remote repair guidance, or schedule on-site services, thereby reducing troubleshooting time and maintenance costs. For connected entertainments, it allows passengers to enjoy online music and videos and interact with users in other vehicles, enhancing travel enjoyment and social experiences, as well as making the driving process more engaging and less monotonous.

\subsection{\textcolor{black}{Lessons Learned}}

\textcolor{black}{This chapter provides the technical background for the combination of the IoV and semantic communication. We discuss the basic architecture and key communication modes of the IoV, as well as the system architecture and types of semantic communication. Through this chapter, we gain an understanding of the communication challenges faced by the IoV and how semantic communication can enhance communication efficiency and security by extracting and transmitting the semantic content of information. In addition, we also explore the potential of integrating the IoV with semantic communication and how this integration can promote the development of intelligent transportation systems.}

\section{Key Technologies of Semantic Communication in IoV}\label{sec3}

Based on the technical background discussed above, the following section will delve into the key technologies of semantic communication in the IoV, including semantic information extraction, communication architecture, and resource allocation, which are critical for realizing efficient and reliable semantic communication.

\begin{figure}[H]
	\hspace{-7.8em}\includegraphics[width=20 cm]{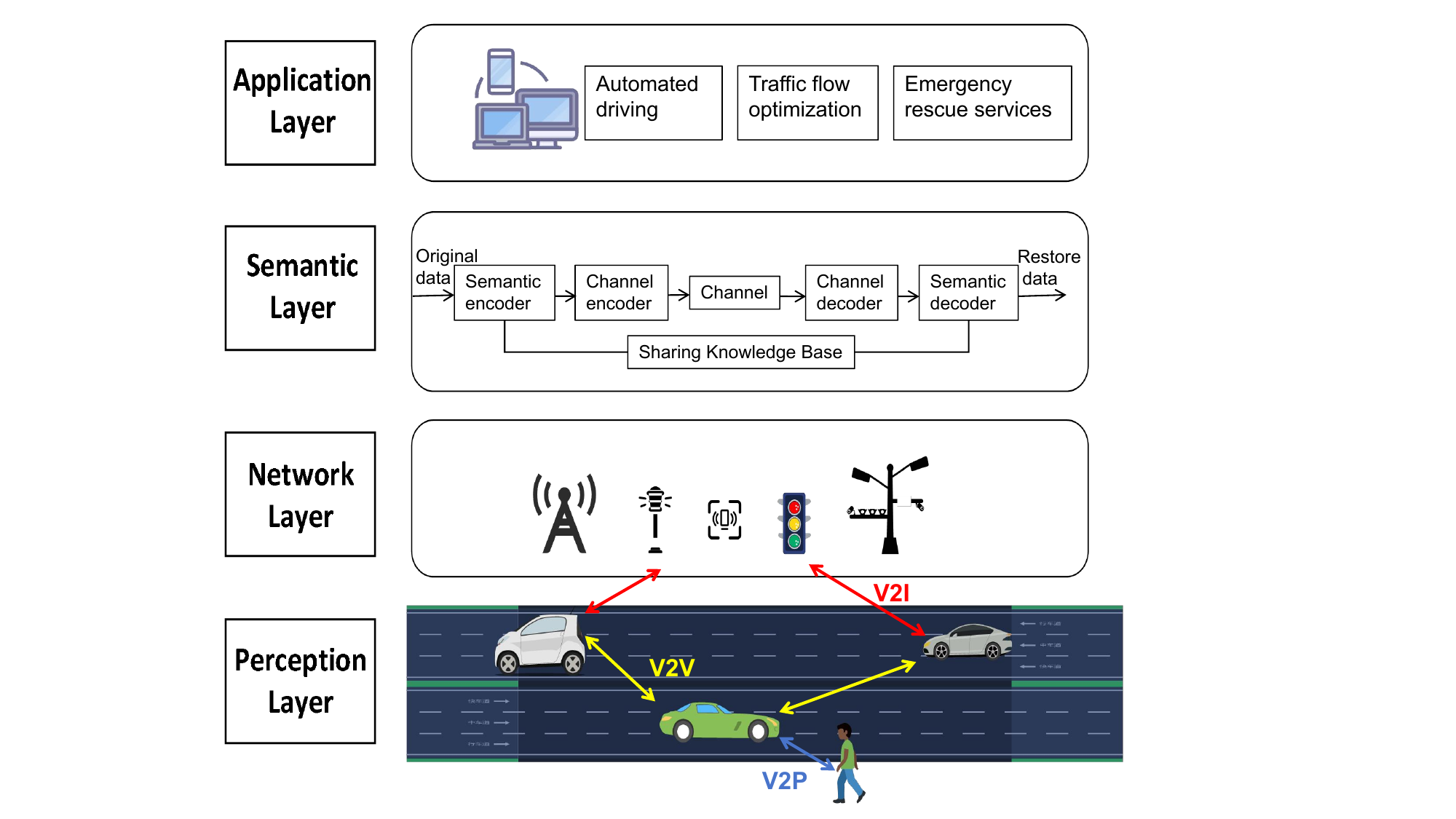}
	\caption{\hl{The} 
 system architecture combining the IoV and semantic communication.\label{fig4}}
\end{figure}

The IoV aims to build an intelligent transportation ecosystem where vehicles, infrastructure, and people are interconnected. This imposes extremely high demands on communication technologies. In the pursuit of higher transmission efficiency, stronger information comprehension capabilities, and more reliable collaborative interactions, semantic communication has emerged as a breakthrough. It goes beyond the traditional signal transmission level of conventional communication, delving into the dimensions of semantic understanding and knowledge sharing. Compared to the traditional three-layer architecture of the IoV, the semantic communication-based IoV introduces an additional semantic layer to facilitate the acquisition and processing of semantic information. For instance, \hl{ref.}
~\cite{9663101}~proposed a four-layer semantic V2X architecture comprising the device layer, network layer, semantic layer, and application layer. The newly added semantic layer extracts key semantic features and compresses redundant information with a semantic encoder and then restores the semantic content via a semantic decoder, ensuring that the receiver accurately understands the sender's intent. The semantic layer also relies on a shared knowledge base to provide a unified foundation for semantic understanding and achieves collaborative optimization between semantic encoding and channel encoding, as well as semantic decoding and channel decoding. This enhances transmission efficiency, reduces data redundancy, overcomes the limitations of traditional communication, and meets the requirements for efficient communication in complex scenarios~\cite{LIU2024528}. Figure~\ref{fig4} shows the system architecture combining the IoV and semantic communication.

\textcolor{black}{To comprehensively evaluate the performance of integrated IoV and semantic communication technologies, we propose a multidimensional unified evaluation framework. This framework covers critical metrics such as semantic accuracy and real-time capability,  enabling systematic quantitative comparisons across methodologies.} These metrics are described as follows:

\begin{itemize}
	\item	Semantic accuracy: This measures the precision of semantic information extraction and reconstruction. It includes general metrics (e.g., BLEU for text, PSNR/SSIM for images, etc.) and task-specific metrics (e.g., IoU for object detection, F1-score for classification, etc.).
	\item	End-to-end latency: This defines the total time from data generation to application reception (encoding + transmission + decoding). This directly impacts real-time-sensitive scenarios like V2V collision warnings.
	\item	Resource efficiency: This includes computational overhead (FLOPs/memory usage) and bandwidth efficiency (data compression ratio), determining feasibility for edge \linebreak  device~deployment.
\end{itemize}

\textcolor{black}{Table ~\ref{tab2} summarizes the complete metric definitions and typical application scenarios. Subsequent sections will analyze the strengths and limitations of each technology based on this framework.}

\begin{table}[H] 
\small
	\caption{\textcolor{black}{A summary of three major categories of evaluation metrics.}\label{tab2}}
	
\begin{adjustwidth}{-\extralength}{0cm}
\centering 
\newcolumntype{C}{>{\centering\arraybackslash}X}
\begin{tabularx}{\fulllength}{cCCC}
		\toprule
		\textbf{Category}	& \textbf{Metric}	& \textbf{Definition}	& \textbf{Applicable Scenarios}\\
		\midrule
		\multirow{9}{*}{Semantic Accuracy}  &  IoU  &  Intersection over Union (overlap between predicted and ground-truth regions)  &   Environmental perception, object detection \\
		\cmidrule{2-4}
		&  mIoU  &  Mean IoU across multiple~classes  &   Complex scene perception \\
		\cmidrule{2-4}
		&  PSNR  &  Peak signal-to-noise ratio (pixel-level difference between original and reconstructed data)  &  Image transmission quality  \\
		\cmidrule{2-4}
		&  F1-score  &  Harmonic mean of precision and recall for classification~tasks  &  Object classification, anomaly~detection  \\
		\midrule
		Real-Time Capability  &  End-to-End Latency	& Total time from data generation to application ({encoding + transmission + decoding}) in ms	& V2V collision alerts, autonomous driving \\
		\midrule
		\multirow{4}{*}{Resource Efficiency}  &  Computational Overhead (FLOPs)  &  Floating-point operations per semantic encoding/decoding~cycle  &   Edge device deployment~feasibility \\
		\cmidrule{2-4}
		&  Bandwidth Efficiency  &  Compression ratio (original data size/transmitted data~size)  &   Image/video transmission \\
		
		\bottomrule
	\end{tabularx}
\end{adjustwidth}
	
\end{table}

\subsection{Semantic Information Extraction}
Semantic information extraction is the foundation of semantic communication and a critical component for efficient information interaction in the IoV. Vehicles need to extract key semantic information from massive sensor data to support real-time perception and intelligent decision making. The development of semantic information extraction technology has evolved from processing single-sensor data to multimodal data fusion. With the rise of deep learning techniques in recent years, the efficiency and accuracy of semantic information extraction have been significantly improved.

In the IoV domain, semantic information extraction predates that of semantic communication. Early studies focused on extracting useful semantic information from single-sensor data or multimodal data (e.g., vision, radar, LiDAR, etc.) to support vehicle perception and decision making. Initial approaches relied on computer vision or multisensor fusion techniques to extract semantic information about the vehicle's surrounding environment (e.g., object detection, road conditions, etc.), thereby enhancing the performance of autonomous driving systems.

\subsubsection{\textcolor{black}{Single-Modal Extraction}}

\textcolor{black}{In the semantic communication system of the IoV, computer vision-based methods dominate the processing of single-modal data. Semantic segmentation techniques are widely used to extract key features, and different methods exhibit distinct advantages. Semantic segmentation is a pixel-level image classification task, aiming to assign each pixel in an image to a predefined category. This process usually relies on deep learning models. The process from image input to semantic information extraction mainly includes the following steps. Firstly, the input image needs to be pre-processed, such as resizing the image and normalizing pixel values, so that the model can process it better. Secondly, feature extraction is performed on the image through convolutional neural networks (such as UNet, DeepLab, etc.). These networks extract high-level semantic features of the image step by step through multiple convolutional operations. For example, the UNet network extracts and fuses image features through the downsampling and upsampling processes. Then, based on the extracted features, the network classifies each pixel to determine its category. This step is usually achieved through a classification layer, such as a fully connected layer or a convolutional layer. Finally, after obtaining the preliminary segmentation result, post-processing is required, such as noise reduction and boundary smoothing, to improve the accuracy and consistency of the segmentation result. The process of semantic information extraction for other data, such as text data, is similar to that of image data, generally involving steps such as text pre-processing, lexical analysis, syntactic analysis, and semantic understanding.}

\textcolor{black}{In the IoV scenario, semantic information extraction technology can be used to extract key information in the road environment, such as vehicles, pedestrians, and traffic signs. For example, \hl{ref.}~\cite{audebert2017segment} employed a deep convolutional model for the semantic segmentation of aerial images and achieved vehicle detection and classification through pixel-level semantic mapping. Their architecture based on SegNet achieved an Intersection over Union (IoU) of 82.4\% and an F1-score of 95.7\% in their vehicle segmentation task. However, the segmentation accuracy of its pyramid pooling module for small targets (such as traffic signs) decreased by approximately 15\%, indicating that this type of method is more suitable for extracting macroscopic objects. \hl{Ref.}~\cite{10.1007/978-981-16-0081-4_63} conducted a comparative study to evaluate the effects of U-Net, Seg-Net, and Fully Convolutional Networks (FCNs) for road segmentation. U-Net achieved a mean Intersection over Union (mIoU) of 94\% on the Camvid dataset, while Seg-Net prioritized computational efficiency in real-time applications. \hl{Ref.}~\cite{9304779} combined semantic segmentation with learned image compression, focusing on the preservation of regions of interest (ROIs). At 0.25 bits per pixel (bpp), their method achieved a peak signal-to-noise ratio (PSNR) of 24.25 decibels (dB) in the ROIs, which is 1.19 dB higher than traditional compression methods. The latest research by \cite{10118717} introduced a multiscale feature extractor based on Swin Transformer, which increased the mIoU by 75\% on the Cityscapes dataset compared to traditional methods. \hl{Ref.}~\cite{10574825} used computer vision algorithms such as RealtimeSeg to extract semantic information from images. Experimental results show that SemCom can significantly increase the load supported by the network, reduce latency, and also perform well in energy consumption.}

\textcolor{black}{From the above methods, it can be seen that different semantic information extraction methods have their own characteristics and advantages. Semantic segmentation methods based on deep convolutional models perform well in vehicle detection and classification tasks and can accurately extract vehicle semantic information. Road detection methods based on U-Net, Seg-Net, and FCNs have their own advantages in distinguishing between road and non-road areas and are suitable for different application scenarios. Methods based on learned image compression can not only extract semantic information but also effectively reduce the amount of data and improve transmission efficiency, especially in enhancing the image quality within the region of interest. The method based on the Swin Transformer has significant advantages in multiscale feature capture and target recognition accuracy in complex traffic scenarios. The semantic segmentation technology in semantic communication performs well in solving the bandwidth bottleneck problem of IoV image transmission and can significantly improve network performance and energy consumption~efficiency.}

\subsubsection{\textcolor{black}{Multimodal Fusion}}
\textcolor{black}{Multimodal fusion integrates heterogeneous sensor data (such as LiDAR, cameras, and Global Positioning System---GPS) to enhance the environmental perception ability. The semantic communication framework further optimizes this process through task-aware feature extraction and transmission.}

\clearpage 
\textcolor{black}{\hl{Ref.}~\cite{8551565} used PSPNet for semantic segmentation and the Random Sample Consensus (RANSAC) algorithm for geometric feature alignment, fusing visual, GPS, and map data. Their Bird's-Eye-View (BEV) registration achieved a feature matching accuracy of over 90\%, effectively eliminating blind spots. \hl{Ref.}~\cite{feng2024semantic} designed a hybrid Convolutional Neural Network--Graph Neural Network--Long Short-Term Memory Network (CNN-GNN-LSTM) encoder to unify multimodal features (images, point clouds, and radar) into a semantic space. This method maintained an F1-score of 0.93 at an SNR of 0 dB, demonstrating strong noise resistance. \hl{Ref.}~\cite{lu2024generative} combined Bird's-Eye-View (BEV) fusion with a Diffusion Model (DM) for noise suppression. At an SNR of 0 dB, the DM-enhanced decoding increased the IoU from 0.65 to 0.80, with further improvements at higher SNR levels.}

\textcolor{black}{In the field of multimodal data fusion, different methods have their unique advantages and applicable scenarios. \hl{Ref.}~\cite{8551565} significantly expanded the vehicle's perception range by integrating multisource data such as visual, GPS, and digital maps, especially excelling in eliminating blind spots with a feature point matching accuracy of over 90\%. The semantic communication in \cite{feng2024semantic} focused on efficiently extracting key semantic information from multimodal data through a semantic encoder. It used multiple neural network architectures to process different modal data, not only significantly reducing the amount of data transmission but also maintaining an F1-score as high as 0.93 under low-SNR conditions. The work of \cite{lu2024generative} integrates multisensor data through technologies such as BEV fusion and uses generative AI technology for noise reduction and enhancement. Its IoU has been significantly improved under different SNR conditions, especially under low-SNR conditions, where the IoU value increased from 0.65 to 0.80, and it can effectively predict future scenarios. Overall, these methods have different emphases in expanding the perception range, reducing data redundancy, enhancing robustness, and improving prediction capabilities, providing diverse solutions for autonomous driving systems.}

\textcolor{black}{From the current research, semantic information extraction technology shows a trend toward more efficient and accurate models. With the increasing requirements for real-time performance and accuracy in the Internet of Vehicles, future research will pay more attention to the lightweight and fast-processing capabilities of models to meet the real-time processing requirements of a large amount of data during high-speed vehicle driving. At the same time, the accuracy and real-time performance of multimodal data fusion will also be a key research direction. Through more advanced fusion algorithms and technologies, the quality and efficiency of semantic information extraction from multimodal data will be further improved.}

\subsection{Semantic Communication Architecture}
In practical application scenarios of the IoV, such as autonomous driving and intelligent traffic coordination management, the realization of complex functionalities not only relies on accurate semantic information extraction but also imposes stringent requirements on the subsequent transmission and interaction modes of information. The design of semantic communication architectures must meet the demands of the IoV for real-time performance, reliability, and efficiency. This section categorizes and summarizes existing semantic communication architectures in the IoV, highlighting their applications and optimization strategies in different scenarios. Table ~\ref{tab3} summarizes the semantic communication architecture in the IoV.

\begin{table}[H]
	\small
	\caption{\hl{Classification} 
 and characteristics of semantic communication architectures in IoV.\label{tab3}}
		\begin{adjustwidth}{-\extralength}{0cm}
				\newcolumntype{C}{>{\centering\arraybackslash}X}
				\begin{tabularx}{\fulllength}{p{4cm} m{4.5cm} X}
				\toprule
				\textbf{Category}	& \textbf{Architecture }	& \textbf{Characteristics}     \\
				\midrule
				\multirow[m]{11}{*}{Multitask and Multiuser}	& Co-SC Architecture~\cite{10013090}			& Multiuser collaboration, with the semantic codec (Sem-Codec) and joint source--channel codec (JSC-Codec) working together to improve semantic reconstruction performance.			\\
				\cmidrule{2-3}
				& Unified Multiuser Semantic Communication System \cite{feng2024semantic}			& Integration of multiuser semantic information, dynamic update of shared and private knowledge bases, and support for joint training of multiple tasks.			\\
				\cmidrule{2-3}
				& Multitask-Oriented Semantic Communication Framework \cite{eldeeb2024multi}			& Multitask processing combining the semantic encoder with task-oriented decoders to support image reconstruction and classification tasks.			\\
				\cmidrule{2-3}
				& SemHARQ Framework~\cite{hu2024semharq}			& Semantic-aware hybrid automatic repeat request (HARQ), feature importance ranking, and distortion evaluation to enhance transmission robustness.			\\
				\midrule
				\multirow[m]{14.5}{*}{Image Transmission}    & ISSC System~\cite{10118717}			& Multiscale semantic feature extraction based on Swin Transformer to reduce data transmission volume while ensuring image quality.			\\
				\cmidrule{2-3}
				& VIS-SemCom System~\cite{lv2024importance}			& Importance-aware image segmentation, multiscale semantic extraction, and importance-aware loss function to improve the segmentation accuracy of important objects.		\\
				\cmidrule{2-3}
				& SEECAD System~\cite{10588546}		& Deep learning-based semantic encoder and decoder, combined with LDPC codes, to transmit semantic information instead of original pixel data.			\\
				\cmidrule{2-3}
				& MTDSC Method~\cite{10473044}			& Multiscene object detection, combined with spatial pyramid pooling and long-short-term memory network to optimize semantic label assignment and transmission.			\\
				\cmidrule{2-3}
				& Combination of Low-latency Routing and Semantic Communication~\cite{10217468}			& Low-latency routing algorithm combined with semantic communication to optimize image data transmission and \linebreak  reduce~latency.			\\
				\midrule

				\multirow[m]{14}{*}{Based on Generative AI}    & AIGC Encoder--Decoder Architecture \cite{feng2023scalable}			& Image-to-text conversion and reinforcement learning optimization to reduce data transmission volume while ensuring \linebreak  image~quality.		\\
				\cmidrule{2-3}
				& Multimodal Semantic-aware Generative AI Framework \cite{10506539}			& Extracts semantic text information and image skeletons and reconstructs images in combination with a generative AI model.			\\
				\cmidrule{2-3}
				& G-MSC Framework~\cite{lu2024generative}			& Multimodal alignment and fusion, with a generative AI-enhanced semantic encoder to improve data quality and communication reliability.			\\
				\cmidrule{2-3}
				& Generative AI-Driven Semantic Communication Framework \cite{raha2023generative}		& Lightweight MSAM extracts key semantic information, combined with GAN to reconstruct images, reducing data transmission volume.			\\
				\cmidrule{2-3}
				& Agent-Driven Generative Semantic Communication Framework (A-GSC) \cite{10815060}			&Generative AI combined with reinforcement learning to dynamically adjust the semantic sampling strategy, improving information interpretability and transmission efficiency.			\\
				\midrule
				\multirow[m]{8}{*}{Other Architectures}    & Blockchain-based Edge-assisted Knowledge Base Management System \cite{LIN2024}			&Blockchain technology ensures data security and consistency, and semantic segmentation reduces data transmission volume. 			\\
				\cmidrule{2-3}
				& PreCMTS Strategy~\cite{10198474}			& Task-driven knowledge-graph-assisted semantic communication; dynamically adjusts semantic unit allocation and \linebreak  relay~selection.			\\
				\cmidrule{2-3}
				& SCKS Framework~\cite{10847825}			& Knowledge sharing of neural network models, with a generative adversarial network (GAN) decoding semantic information to update the vehicle model.			\\
				\bottomrule
			\end{tabularx}
			\end{adjustwidth}
	\end{table}

\subsubsection{Multiuser Collaboration and Multitask Driving} 
In the IoV, multiuser collaboration and multitask driving are two closely related semantic communication scenarios. Multiuser collaboration focuses on semantic information sharing between vehicles, while multitask driving focuses on optimizing the processing and transmission of semantic information for different tasks.

\textcolor{black}{\hl{Ref.}~\cite{10013090} proposed a multiuser collaborative semantic communication architecture (Co-SC), which comprises components such as semantic codecs (Sem-Codec), joint source--channel codecs (JSC-Codec), and task-related modules. The Co-SC architecture significantly enhances semantic reconstruction performance and recognition accuracy under low-SNR conditions. By leveraging multiuser collaboration, Co-SC effectively exploits the semantic correlation among users, reducing the transmission of redundant information and thereby significantly improving communication efficiency. However, the performance improvement of this architecture is limited under high-SNR conditions, and it has a high dependence on channel state information (CSI), requiring accurate channel estimation. Moreover, the performance of the architecture is highly reliant on the pre-trained knowledge base and offline-trained models, and its adaptability to the dynamically changing VANET environment needs further investigation. Overall, the Co-SC architecture provides an effective solution for semantic communication in VANETs, but a trade-off between computational resources and real-time requirements is necessary in practical deployment.}

\textcolor{black}{\hl{Ref.}~\cite{feng2024semantic} presented a unified multiuser semantic communication system that integrates semantic information from multiple users to achieve collaborative processing. This framework significantly reduces the amount of data transmission through semantic encoding. For instance, in the task of object classification, it achieved performance comparable to or even better than traditional methods with a compression rate of only about 8.4\%. The F1-score reached 0.9880 under high-SNR conditions and remained above 0.93 under low-SNR conditions, demonstrating good robustness. The system not only supports multitask processing but also ensures semantic understanding and vehicle privacy by constructing shared and private knowledge bases. However, the system complexity is relatively high, requiring substantial computational resources to support joint training of multiple tasks, which may pose a challenge to the real-time processing capabilities of vehicles in VANETs. Moreover, the adaptability of the system under dynamic channel conditions needs further~improvement.}

\textcolor{black}{\hl{Ref.}~\cite{eldeeb2024multi} proposed a multitask semantic communication framework for autonomous vehicles. It employs convolutional autoencoders for semantic encoding of traffic sign images and utilizes satellite communication for information sharing among vehicles. The image reconstruction quality is assessed using the structural similarity index measure (SSIM), and the results show that it outperformed the traditional 16-QAM scheme under low-SNR conditions, indicating good performance in semantic similarity. Additionally, the framework significantly improved data transmission efficiency by reducing the amount of transmitted data by up to 89\%, indirectly reducing transmission latency. However, the paper does not provide a detailed discussion on computational overhead. Considering its implementation based on deep learning, it may face certain computational resource demands in real-time deployment.}

\textcolor{black}{\hl{Ref.}~\cite{hu2024semharq} introduced a semantic-aware hybrid automatic repeat request (SemHARQ) framework for multitask semantic communication in VANETs. This framework significantly enhances the efficiency and robustness of semantic feature transmission through feature importance ranking (FIR) and feature distortion evaluation (FDE) networks. The study demonstrates excellent performance under low-SNR conditions. For example, in the task of vehicle reidentification, the rank-1 accuracy was improved by more than 20\% compared to existing methods, and the vehicle color classification accuracy was increased by 10\%. However, despite its superior performance in multitask scenarios, the computational overhead may limit its real-time deployment in VANETs, especially on resource-constrained edge devices. Moreover, the study mainly focuses on the transmission efficiency and accuracy of semantic features, with less discussion on key indicators for practical deployment such as transmission delay and computational overhead. Therefore, although SemHARQ has significant theoretical advantages, further optimization is needed in practical applications to reduce computational complexity and meet real-time requirements.}

\textcolor{black}{In the research on multiuser collaboration and multitask-driven semantic communication frameworks, these studies have proposed various distinctive solutions. The Co-SC architecture significantly improves semantic reconstruction performance through multiuser collaboration under low-SNR conditions but has a high dependence on CSI and limited performance improvement under high-SNR conditions. The system in \cite{feng2024semantic} significantly reduces data transmission through semantic encoding and supports multitask processing, but it has high system complexity and a large demand for computational resources. The framework in \cite{eldeeb2024multi} improves image reconstruction quality and transmission efficiency through semantic encoding in the autonomous driving scenario, but the issue of computational overhead needs to be clarified. The SemHARQ framework significantly enhances the efficiency and accuracy of semantic feature transmission through feature importance ranking under low SNR conditions, but its computational overhead may limit real-time deployment. These studies each have their own advantages, but in practical VANET applications, it is necessary to comprehensively consider factors such as computational resources, real-time performance, and adaptability to channel conditions to achieve more efficient and practical semantic communication solutions. Future work may explore the integration of the advantages of these methods to optimize computational overhead and real-time performance, thereby further enhancing the overall performance of semantic communication in VANETs.}

\subsubsection{Oriented to Image Transmission}
Image transmission is one of core tasks in the IoV for environmental perception and decision making. Efficient image transmission requires not only reducing data volume but also ensuring that image quality meets the real-time demands of autonomous driving and intelligent transportation systems. For specific tasks such as image segmentation and object detection, researchers have proposed various optimized semantic communication architectures to address the needs of different application scenarios.

\textcolor{black}{\hl{Ref.}~\cite{10118717} proposed an Image Segmentation Semantic Communication (ISSC) system for the IoV. This system utilizes a multiscale semantic feature extractor and a semantic feature aggregator based on the Swin Transformer to achieve an efficient transformation from the input image to semantic features. Additionally, it accurately reconstructs the image segmentation at the receiving end through a semantic feature decoder and a reconstructor. Experimental results show that the ISSC system performs excellently in a low-SNR environment, with its mean Intersection over Union (mIoU) increased by 75\% compared to traditional coding methods, indicating that the system has a significant advantage in semantic similarity. Furthermore, although the paper does not explicitly mention the specific values of transmission delay and computational overhead, considering its deep learning-based architecture, the computational overhead may be high, especially in real-time deployment scenarios. }

\textcolor{black}{\hl{Ref.}~\cite{lv2024importance} proposed a semantic communication system for autonomous driving based on importance-aware image segmentation (VIS-SemCom). Through multiscale semantic feature extraction and an importance-aware loss function, it significantly improves the accuracy of image segmentation and communication efficiency. Experiments show that under the requirement of an average mIoU of 60\%, this system can achieve a coding gain of nearly 6 dB compared to traditional solutions and performs well under low-SNR conditions, with the maximum reduction of the transmitted data volume reaching 70\%. The system improved the segmentation accuracy of key objects (such as vehicles and pedestrians) by approximately 4\%, but its generalization ability in complex environments still needs to be further verified.}

\textcolor{black}{\hl{Ref.}~\cite{10588546} proposed a semantic end-to-end communication system named SEECAD, which is specifically designed for image transmission in autonomous driving scenarios. By constructing a semantic encoder and decoder through a deep learning architecture and combining with Low-Density Parity-Check (LDPC) codes, it efficiently transmits the semantic information of images instead of the original pixel data, thus significantly reducing the amount of transmitted data and improving the transmission efficiency. At the receiving end, the system reconstructs the image segmentation mask using a shared knowledge base, enhancing its robustness in noisy environments. Experimental results show that under 4QAM modulation and LDPC (16,32) coding, it achieved a segmentation accuracy of 97.4\% and an IOU value of 0.75 at an SNR of 12 dB, demonstrating high semantic similarity and anti-noise ability. However, the training and real-time inference of the deep learning model require high computational resources, which may limit its real-time deployment in the IoV.}

\textcolor{black}{\hl{Ref.}~\cite{10473044} proposed a semantic communication method for multiscene object detection in intelligent vehicle networks (MTDSC), aiming to improve the processing efficiency of image data through semantic encoding and transmission. This method uses a convolutional neural network and a region proposal network for object detection, combines Spatial Pyramid Pooling (SPP) with a long short-term memory network (LSTM) for semantic label assignment, and finally achieves reliable transmission through a variational autoencoder. In terms of model performance evaluation, MTDSC performs well in semantic similarity. Experiments have verified its high mIoU values in different road scenes (such as 83.276\% in highway scenes), indicating that it can accurately understand and transmit key semantic information. In terms of transmission delay, this method reduces the amount of data through efficient semantic encoding, thus reducing the communication delay. However, its limitation lies in the high computational overhead, especially in the training and inference stages of the deep learning model, which may pose a challenge to the real-time deployment in the IoV.}

\textcolor{black}{\hl{Ref.}~\cite{10217468} combined low-latency routing and semantic communication to optimize the transmission of image data in the IoV. In this method, the low-latency routing algorithm takes into account factors such as the distance between vehicle nodes, the packet loss probability of candidate nodes, remaining energy, and load, and it selects the optimal routing node to ensure the rapid transmission of image data. At the same time, semantic communication technology uses a convolutional neural network to extract image features, obtains a semantic sequence through a feature-semantic transformation function, compresses the feature map according to the task importance, and then transmits it. The receiving end performs the inverse operation to restore the image for classification and recognition. This method mainly addresses the problem of high transmission latency of image data in the IoV caused by the high-speed movement of vehicles and the dynamic changes of the network topology. Experiments have proven that compared with traditional methods, this method can significantly reduce the transmission latency and has little impact on the accuracy of image classification. Its advantage lies in significantly reducing the transmission latency and maintaining a high image classification accuracy. However, the high computational overhead and the lack of detailed evaluation limit its feasibility for real-time deployment in resource-constrained environments.}

\textcolor{black}{In the research on semantic communication for image transmission in the IoV, different methods have shown their unique advantages and limitations. The ISSC system demonstrates excellent semantic similarity in a low-SNR environment through efficient semantic feature extraction and transformation, with a significant increase in the average mIoU. However, the high computational overhead brought by the deep learning architecture may restrict real-time deployment. The VIS-SemCom system performs excellently in terms of image segmentation accuracy and communication efficiency, with obvious coding gain and a significant reduction in the transmitted data volume. However, its generalization ability in complex environments, as well as the detailed information about computational overhead and transmission delay, remains unclear. The SEECAD system has excellent semantic similarity and anti-noise ability at high SNRs, but it has a certain dependence on the SNR, and the training and inference of the deep learning model require high computational resources. The MTDSC method performs well in semantic similarity for multiscene object detection and reduces the consumption of computational resources through reinforcement learning optimization, but the computational overhead is still high. Combining low-latency routing and semantic communication to optimize image data transmission significantly reduces the transmission latency and maintains a high image classification accuracy, but it also faces the problems of high computational overhead and lack of detailed evaluation. These methods have their own focuses in improving semantic similarity, reducing the transmitted data volume, and decreasing the transmission latency, but there are still challenges to varying degrees in terms of computational overhead, real-time performance, generalization ability, and adaptability to the SNR. Further research is needed in the future to balance these factors to better meet the actual needs of image transmission in the IoV.}

\subsubsection{Generative AI-Based Semantic Communication Architectures}
Generative AI technologies, such as generative adversarial networks (GANs) and large language models (LLMs), have shown significant potential in optimizing data processing and transmission efficiency in semantic communication, particularly in the context of the IoV. By transforming complex data (e.g., images, text, etc.) into concise semantic representations and leveraging multimodal fusion and reconstruction techniques, generative AI substantially reduces data volume while enhancing the intelligence and accuracy of communication. Current generative AI-based semantic communication architectures in the IoV primarily focus on image transmission. Unlike traditional image transmission architectures, these approaches emphasize the use of generative AI to intelligently process complex data. By converting images, text, and other data types into compact semantic representations and integrating multimodal fusion and reconstruction, these architectures not only reduce data transmission but also significantly improve communication intelligence, thus offering new perspectives and methods for semantic communication in the IoV. Below are several generative AI-based semantic communication architectures and their applications.

\textcolor{black}{\hl{Ref.}~\cite{feng2023scalable} proposed an scalable Artificial Intelligence Generated Content (AIGC) encoder--decoder architecture. The encoder leverages large language models to convert images into concise text prompts. Before transmission, the text information is optimized by a reinforcement learning module. The decoder then converts the optimized text back into images and decides whether to transmit semantically important image regions based on bandwidth conditions. The system demonstrates excellent performance in semantic similarity. Experiments have validated its advantages in image reconstruction quality while significantly reducing the amount of transmitted data, with a compression ratio of several times. This is highly beneficial for bandwidth-constrained VANET environments.}

\textcolor{black}{\hl{Ref.}~\cite{10506539} proposed a multimodal semantic-aware framework based on generative AI. It achieves efficient data transmission and reconstruction by extracting semantic text information and image skeletons. The framework first extracts semantic text information and image skeletons from real-time road images and then combines them into small data packets for V2V communication. The receiving end uses a generative AI model to generate road condition images that match reality and reconstructs them for display to intelligent assistants. This significantly reduces data volume and effectively reduces transmission delay. However, although the framework emphasizes data privacy protection, it does not provide detailed explanations on how to ensure data security in practical deployment. Therefore, while the framework excels in improving semantic similarity and transmission efficiency, it still needs further optimization in terms of computational overhead and privacy security.}

\textcolor{black}{\hl{Ref.}~\cite{lu2024generative} proposed a Generative AI-Enhanced Multimodal Semantic Communication (G-MSC) framework enhanced by generative AI. It reduces data redundancy through multimodal alignment and fusion and performs denoising and semantic inference on noisy data at the receiver. In the G-MSC framework, generative AI technology enhances the capabilities of the semantic encoder, enabling it to better handle multimodal data. The optimization of channel transmission and the semantic decoder also improves the reliability and stability of communication. Experimental results show that diffusion models significantly improve image clarity and IoU. Particularly under low-SNR conditions, it effectively mitigates the impact of channel noise on performance. The framework's strengths lie in its efficient processing of multimodal data and adaptability to dynamic environments. However, its limitation is the high demand for computational resources, which may restrict its real-time deployment in resource-constrained VANET environments. Future research directions include hybrid digital--analog transmission, multivehicle semantic information scheduling, and cross-task coordination, which are expected to further optimize the performance and application scope of the G-MSC framework.}

\textcolor{black}{\hl{Ref.}~\cite{raha2023generative} proposed a GAI-driven semantic communication framework for next-generation wireless networks (such as 6G). The framework extracts key semantic information from images using the lightweight Mobile Segment Anything Model (MSAM) and reconstructs and denoises images at the receiver using a generative adversarial network (GAN), achieving high-quality image reconstruction under different SNR conditions. Experimental results show that models trained under specific SNR conditions can achieve high PSNRs. Moreover, the framework further reduces data transmission by periodically updating background information. Experimental results indicate that the framework achieved a significant reduction in data transmission volume, up to 93.45\%, while maintaining the integrity of the original content. However, the framework is sensitive to the SNR during training, with limited robustness. The real-time and feasibility of its deployment in VANETs still need further investigation.}

\textcolor{black}{\hl{Ref.}~\cite{10815060} proposed an Agent-Driven Generative Semantic Communication (A-GSC) framework based on reinforcement learning. It seamlessly integrates semantic extraction and semantic sampling using generative AI. By transmitting the semantic information of a scene in text form, it reduces data volume while improving information interpretability. Moreover, the semantic sampling agent based on reinforcement learning can dynamically adjust sampling strategies according to channel conditions and changes in source data, ensuring optimal semantic information transmission under limited energy consumption. The framework performs well in terms of semantic similarity, transmission delay, and computational overhead. Its strengths are high semantic similarity, low transmission delay, and small computational overhead, making it suitable for real-time applications in VANETs. However, its limitations include insufficient adaptability to complex communication environments, and the feasibility of real-time deployment on resource-constrained devices still needs further verification.}

\textcolor{black}{In summary, these generative AI-based semantic communication frameworks each have their own characteristics. The AIGC architecture excels in data compression but lacks sufficient evaluation of real-time performance and computational overhead, with its deployment feasibility remaining to be verified. The multimodal framework has high data transmission efficiency but demands substantial computational resources and lacks clear privacy security guarantees. The G-MSC framework has strong capabilities in processing multimodal data and adapting to dynamic environments, but its high demand for computational resources restricts real-time deployment. The framework in \cite{raha2023generative} achieves high semantic similarity and significant reduction in data transmission under specific SNR conditions, but it is sensitive to the SNR, with limited robustness, and the real-time and feasibility of its deployment in VANETs need further investigation. The A-GSC framework performs well in semantic similarity, transmission delay, and computational overhead, making it suitable for real-time applications in VANETs, but it has insufficient adaptability to complex communication environments, and the feasibility of real-time deployment on resource-constrained devices remains to be verified.}

\textcolor{black}{Overall, these methods have their own advantages in terms of data transmission efficiency and semantic similarity, but they also have varying degrees of shortcomings in computational overhead, real-time performance, robustness, privacy security, and deployment feasibility. Future research needs to balance and optimize these aspects to better meet the practical needs of integrating VANETs and semantic communication.}

\subsubsection{Other Architectures}
In addition to the mainstream semantic communication architectures mentioned above, several innovative architectures have demonstrated unique advantages in the IoV. These architectures expand the application scenarios of semantic communication and enhance communication efficiency and intelligence by integrating technologies such as blockchain, knowledge graphs, and neural network sharing.

The fusion of blockchain and edge computing technologies brings multiple benefits to semantic communication, including data security, privacy protection, and improved system efficiency. \hl{Ref.}~\textcolor{black}{\cite{LIN2024} proposed a blockchain-based edge-assisted knowledge base management system for semantic communication in the IoV. This system extracts key information through semantic segmentation and utilizes blockchain sharding technology to enhance the efficiency and security of knowledge base management. Experiments show that the proposed semantic communication method has lower transmission latency compared with traditional communication methods under low-SNR conditions and performs better in terms of transmission efficiency. Its advantages lie in effectively reducing communication costs, improving data transmission efficiency, and ensuring the consistency and security of the knowledge base with the help of blockchain technology. However, the limitation of this study is that it does not discuss in detail the computational overhead in the real-time deployment of the IoV, as well as the scalability issues in large-scale networks, which may affect its widespread application in practical scenarios.}

\hl{Ref.}~\cite{10198474} proposed a task-driven semantic-aware green cooperative transmission strategy (PreCMTS) that is suitable for intermittently connected IoV scenarios. This strategy employs a weighted directed graph to achieve semantic-aware transmission by analyzing parameters such as the remaining dwell time of vehicles at RSUs, encounter time with target vehicles, and lifetime of V2V links. It derives an expression for achievable throughput that meets delay requirements and formulates the coupled problem of semantic unit allocation and predictive relay selection as a combinatorial optimization problem. A low-complexity algorithm based on Markov approximation is designed to solve this problem. PreCMTS primarily addresses the intermittent connectivity caused by the high deployment costs and energy consumption of the infrastructure in vehicular networks, as well as the limited applicability of existing semantic communication research in dynamic and complex vehicle networks. The strategy aims to achieve both semantic and green communication. Experimental results show that PreCMTS effectively reduces energy consumption, improves the reliability of semantic transmission, and enhances semantic energy efficiency, thus outperforming baseline methods under various conditions.\textcolor{black}{However, this strategy has deficiencies in terms of the analysis of computational overhead and feasibility for real-time deployment, and it fails to thoroughly explore the computational complexity and resource consumption of the deep learning model in the IoV environment.  }

In the integration of the IoV and semantic communication, besides transmitting IoV data through semantic communication systems, neural network models can also be shared. \hl{Ref.}~\cite{10847825} proposed a deep semantic communication framework, SCKS, to address the efficiency issues of knowledge sharing of neural network models in the IoV. This architecture extracts semantic features of neural network models using a dataset distillation algorithm and designs a semantic decoding algorithm based on GANs. By leveraging the deep semantic communication system, RSUs extract semantic information from neural network models, encode it into semantic feature vectors, and transmit it to vehicles. Upon receiving the vectors, vehicles decode them into synthetic datasets using a semantic decoder and then update their neural network models. This enables efficient knowledge sharing from RSUs to multiple vehicles. \textcolor{black}{The experimental results demonstrate that the framework exhibits remarkable performance in semantic similarity, effectively extracting the semantic features of neural network (NN) models. Under low-SNR conditions, the framework outperformed traditional methods in terms of transmission delay. However, the framework incurs a high computational overhead, particularly during the training phase, which poses limitations on the feasibility of real-time deployment. Despite these challenges, the framework, referred to as SCKS, performs well on high-resolution datasets. Further investigation is required to assess its performance when the number of categories increases.}

These innovative architectures, through task-driven approaches, knowledge graph optimization, and neural network sharing, further expand the application scenarios of semantic communication in the IoV. \textcolor{black}{However, they also demonstrate their respective advantages and limitations. The blockchain-based edge-assisted knowledge base management system performs excellently in terms of data security and transmission efficiency. However, the issues of computational overhead in the real-time deployment of the IoV and scalability in large-scale networks have not been resolved yet. The PreCMTS effectively reduces energy consumption and improves the reliability of semantic transmission, but it lacks a detailed analysis of the computational complexity and resource consumption of the deep learning model. The deep semantic communication framework SCKS enhances semantic similarity and transmission efficiency through the knowledge sharing of the neural network model. However, the relatively high computational overhead, especially during the training phase, limits its feasibility for real-time deployment. Although these architectures are innovative in their own ways, in practical applications, various factors such as computational overhead, scalability, and energy consumption need to be comprehensively considered to better adapt to the complex and changeable environmental requirements of the IoV and achieve efficient and reliable semantic communication. }


\subsection{Resource Allocation and Management}
In intelligent networking scenarios, semantic-oriented resource allocation methods have significant advantages~\cite{Chen2022}. Similarly, in the context of the IoV, the high mobility of vehicles and the diverse semantic communication tasks make resource allocation and management critically important. Different tasks have time-varying demands for resources such as spectrum, power, and computing. For instance, road condition information transmission requires low latency and high bandwidth, while vehicle diagnostics emphasize data accuracy and completeness. This subsection reviews relevant research progress and categorizes resource allocation and management strategies based on different technical approaches. Table~\ref{tab4} summarizes existing research on resource allocation and management in the IoV based on semantic communication.

\begin{table}[H]
\small
	\caption{\hl{Resource} 
 allocation and management methods in IoV based on semantic communication.\label{tab4}}
		\begin{adjustwidth}{-\extralength}{0cm}
			\newcolumntype{C}{>{\centering\arraybackslash}X}
				\begin{tabularx}{\fulllength}{cCCC}
				\toprule
				\textbf{Category}	& \textbf{Methodology}	& \textbf{Characteristics}     & \textbf{Application scenario}\\
				\midrule
				\multirow[m]{20}{*}{Reinforcement Learning}	& SARADC Framework~\cite{10636300}			& Significantly improves HSSE and semantic throughput, adapting to high-resolution and low-signal-to-noise-ratio~conditions.			& 5G-V2X heterogeneous networks \\
				\cmidrule{2-4}
				& SSS Algorithm~\cite{10644092}			& Dynamically optimizes spectrum sharing, improving semantic transmission efficiency and spectrum utilization.			& Spectrum sharing in vehicle networking\\
				\cmidrule{2-4}
				& DDQN Method~\cite{10506539}			& Ensures efficient semantic information transmission and supports multiagent collaborative resource~allocation.			& Generative AI-empowered V2V communication in IoV\\
				\cmidrule{2-4}
				& SAMRA Algorithm~\cite{shao2024semantic3}			& Adapts to multiscenario changes, improving resource allocation efficiency and task success~rate.			& C-V2X platoon communication\\
				\cmidrule{2-4}
				& VSRAA-SM Algorithm~\cite{9538906}			& Improves video semantic understanding accuracy, reducing the CUE outage probability and V2V transmission~rate.			& In-vehicle video semantic resource allocation\\
				\midrule
				\multirow[m]{10}{*}{Optimization Theory}    & Lyapunov Optimization Method~\cite{10073623}			& Features fast convergence, improving system robustness and power consumption efficiency.			& D2D in-vehicle networks\\
				\cmidrule{2-4}
				& Two-stage Suboptimal Solution Method~\cite{10681786}			& Maximizes semantic detection accuracy, minimizes wireless resource costs, and ensures communication link quality.			& Vehicle platoon collaboration\\
				\cmidrule{2-4}
				& Two-stage Stochastic Integer Programming (SSTS)~\cite{10570867}		& Optimizes resource allocation, reduces transmission costs, and supports immersive~experiences.			& Virtual transportation network in the meta-verse\\
				\midrule
				\multirow[m]{7}{*}{Federated Learning}    & MSFTL Framework~\cite{10416926}			& Reduces vehicle computational costs, improves resource utilization efficiency, and supports distributed~training.			& In-vehicle semantic communication\\
				\cmidrule{2-4}
				& FVSCom Framework~\cite{10333738}			& Improves computational efficiency and semantic extraction accuracy, enhancing robustness to vehicle departure or~withdrawal.			& In-vehicle semantic communication\\
				\bottomrule
			\end{tabularx}
			\end{adjustwidth}
	\end{table}

\subsubsection{Reinforcement Learning-Based Resource Allocation Methods}

\textcolor{black}{In the context of integrating VANETs with semantic communication, reinforcement learning (RL)-based resource allocation methods have garnered significant attention due to their strong dynamic adaptability and optimization capabilities. These methods can dynamically adjust resource allocation strategies in real time according to the dynamic environment and task requirements in VANETs, thereby significantly enhancing \linebreak  system performance.}

\textcolor{black}{\hl{Ref.}~\cite{10636300} proposed a Semantic-Aware Resource Allocation and Decision-making Framework (SARADC) specifically for the complex scenario of spectrum sharing between vehicles and Wi-Fi users in 5G-V2X heterogeneous networks. This research introduces innovative metrics such as high-speed semantic transmission rate (HSR) and high-speed semantic spectrum efficiency (HSSE). The near-policy optimization (PPO) algorithm is employed to optimize parameters, including channel allocation, power allocation, duty cycle, and semantic symbol length. Experimental results demonstrate that the framework significantly improves the HSSE and semantic throughput (ST) under high-resolution and low-SNR conditions, effectively enhancing resource utilization efficiency and system performance. However, the discussion on computational overhead and transmission delay is insufficient, which may limit its real-time deployment in VANETs.}

\textcolor{black}{Similarly, \hl{ref.}~\cite{10644092} proposed a Semantic-Aware Spectrum Sharing (SSS) algorithm based on Deep Reinforcement Learning (DRL) to address the spectrum sharing problem in the IoV. This algorithm redefines spectrum sharing metrics such as HSSE and the HSR by incorporating semantic information and utilizes the soft actor-critic (SAC) method to optimize decisions to maximize HSSE and improve the success rate of effective semantic information transmission (SRS). Simulation results show that the SSS algorithm outperforms traditional bit-based spectrum sharing methods in terms of HSSE and SRS. However, the limitation of this algorithm is its potentially high computational overhead, especially considering the computational capabilities and resource constraints of vehicles in real-time deployment.}

\textcolor{black}{\hl{Ref.}~\cite{10506539} started from the perspective of generative AI and used a deep reinforcement learning (DRL) algorithm to design a resource allocation strategy for V2V communication in the IoV. The study uses the double deep Q-network (DDQN) method to optimize parameters such as channel selection, transmission power, and diffusion steps, ensuring efficient semantic information transmission in VANETs. Additionally, the method designs action space, state space, and reward functions adapted to semantic communication to accommodate the dynamically changing channel conditions and transmission requirements in VANETs. Its advantage lies in dynamically adjusting resource allocation using DRL to adapt to the dynamics of vehicular networks, effectively reducing transmission delay and improving semantic similarity. However, the study lacks in-depth discussion on the quantification of computational overhead and lacks discussion on the weight allocation of multimodal data fusion, which may affect its robustness in complex scenarios and scalability of practical applications.}

\textcolor{black}{\hl{Ref.}~\cite{shao2024semantic3} aimed at the difficult problem of C-V2X platoon communication resource management and proposed a Semantic-Aware Multimodal Resource Allocation (SAMRA) algorithm using Multiagent Reinforcement Learning (MARL). The study defines metrics and quality of experience (QoE) concepts suitable for semantic and multimodal data in the system model, with maximizing the QoE and the success rate of V2V semantic information transmission as joint optimization objectives. By optimizing channel allocation, power allocation, and semantic symbol length using MARL, experimental results show that the algorithm outperformed baseline methods in multiple scenarios. Its advantage is the enhanced scalability and adaptability of the system through distributed decision making, effectively improving the QoE and the success rate of semantic information transmission (SRS). However, similar to the previous studies, this research lacks in-depth discussion on computational overhead in real-time deployment and lacks quantitative comparison among different methods, making it difficult to comprehensively assess its feasibility in practical VANET scenarios.}

\textcolor{black}{Finally, \hl{ref.}~\cite{9538906} aimed at the dilemma of in-vehicle video semantic resource allocation and built a model based on the multiagent deep Q-network (MADQN) for optimization according to the tasks between vehicles, base stations, and vehicles. Experimental results show that compared to traditional algorithms, the VSRAA-SM achieved higher accuracy in video semantic understanding under different vehicle transmission power and bandwidth conditions and also performed better in terms of the CUE outage probability and V2V transmission rate, making it more suitable for spectrum reuse scenarios. However, the limitation of this study is the lack of detailed analysis on computational overhead, with the feasibility of real-time deployment remaining unclear. Moreover, the paper does not construct a unified evaluation framework, posing certain difficulties for quantitative comparison with other methods.}

\textcolor{black}{In summary, these RL-based resource allocation methods have demonstrated significant advantages in enhancing semantic communication performance. However, they still have shortcomings in terms of computational overhead, the feasibility of real-time deployment, and multimodal data fusion. Future research needs to further optimize these methods to reduce computational overhead and improve their adaptability and scalability in practical VANET scenarios. Additionally, constructing a unified evaluation framework to quantitatively compare the performance of different methods will be crucial for advancing this~field.}

\subsubsection{ Optimization Theory-Based Resource Allocation Methods}
Optimization algorithms are also widely used in resource allocation to enable efficient resource distribution through mathematical modeling and optimization techniques. Compared to reinforcement learning methods, optimization algorithms often provide more precise solutions for resource allocation problems under specific constraints, demonstrating superior performance in system stability and robustness. For example, \hl{ref.}~\cite{wang2020stackelberg} proposed an RSU caching incentive scheme based on the Stackelberg game and optimized the content caching strategy and pricing mechanism by establishing a game model between the base station (BS) and RSU. In the field of the combination of vehicle networking and semantic communication, researchers have also proposed some resource allocation schemes based on optimization theory, significantly improving the efficiency and performance of semantic communication in scenarios such as Device-to-Device (D2D) communication and vehicle platooning in the IoV.

\hl{Ref.}~\cite{10073623} proposed a long-term robust resource allocation scheme for D2D in-vehicle networks. The scheme also considers the semantic access control of the application layer and the power control of the physical layer. The Lyapunov optimization method is used to transform the long-term constraints into queue stability conditions. Then, successive convex approximation and the Karush--Kuhn--Tucker (KKT) condition are used to solve the subproblem, while Bernstein approximation is used to deal with uncertain probability constraints. Simulation results show that the algorithm enables rapid convergence of metrics such as the semantic access rate and power. Under different interruption probability thresholds, the algorithm demonstrates the trade-off between the transmission rate and delay. Compared to traditional methods, it exhibited superior performance in system robustness and power consumption. \textcolor{black}{Simulation results show that this algorithm performs excellently in terms of system robustness and power consumption. However, there is relatively little discussion on the computational overhead and real-time performance in practical deployment, which limits the feasibility of its application in an IoV environment with limited~resources.}

Beyond resource allocation in D2D vehicular networks, optimization theory has also been extensively applied in vehicle platooning scenarios. \hl{Ref.}~\cite{10681786} proposed an innovative resource allocation scheme based on semantic communication for the collaborative data processing scenario between autonomous vehicle platoons and base stations in the IoV. The scheme aims to maximize semantic detection accuracy and minimize wireless resource costs while incorporating constraints such as bandwidth, transmission power, and SNR. Given the NP-hard nature of the optimization problem, a two-stage suboptimal solution is adopted. In the first stage, a bipartite graph is used to transform the communication mode selection problem into a maximum weight matching problem, and the Hungarian algorithm is applied to determine the matching relationship between vehicles and task processing units, ensuring communication link quality and reliability. In the second stage, based on the results of the first stage, the resource allocation problem is transformed into a convex optimization problem, which is solved using Matlab's fmincon function to achieve reasonable allocation of resources such as bandwidth and transmission power. {Simulation results show that, compared to the scheme that only used V2I communication, this method significantly improved the accuracy of video semantic detection, especially showing obvious advantages when the distance between vehicles is large. However, the paper does not provide specific data on key indicators such as semantic similarity, transmission delay, and computational overhead, which limits the comprehensive evaluation of the performance of the scheme.}

Additionally, \hl{ref.}~\cite{10570867} proposed a Stochastic Semantic Transmission Scheme (SSTS) based on two-stage Stochastic Integer Programming (SIP) to address resource allocation problems in virtual traffic networks within the metaverse. By introducing semantic communication technology, the scheme leverages the sensing capabilities of edge devices to transmit data from the physical world to the virtual world, providing immersive experiences for Virtual Service Providers (VSPs). The SSTS scheme considers the demand uncertainty of VSPs and optimizes resource allocation through a combination of reservation and on-demand plans, reducing transmission costs. {Experiments have verified the advantages of the SSTS in energy consumption optimization. The energy consumption of semantic data transmission is significantly lower than that of non-semantic data transmission, which indicates that it has significant advantages in reducing energy consumption. However, the limitation of this study is that it mainly focuses on the transmission cost of semantic data and the subscription strategy, and the analysis of the computational overhead and complexity in real-time deployment is not in-depth enough.}

\textcolor{black}{In these studies on resource allocation for semantic communication in the IoV, methods based on optimization theory exhibit diverse characteristics and advantages, but they also have their own limitations. The D2D in-vehicle network resource allocation scheme proposed in reference \cite{10073623} uses methods such as Lyapunov optimization, which significantly enhances the system's robustness in high-speed mobile scenarios. However, it lacks sufficient discussion on computational overhead and real-time performance. In \mbox{reference \cite{10681786}}, for the scenario of vehicle platoon collaboration, technologies such as bipartite graph matching are employed, effectively improving the accuracy of semantic detection. Nevertheless, there is a lack of comprehensive evaluation of key indicators such as semantic similarity and transmission delay. The SSTS scheme, on the other hand, focuses on energy consumption optimization and demonstrates the advantage of semantic data transmission in reducing energy consumption. But the analysis of computational overhead and complexity in real-time deployment is not in-depth enough.}

\textcolor{black}{These methods have different focuses in terms of objectives, technical means, and application scenarios. However, in practical applications, various factors such as system stability, resource utilization efficiency, real-time performance, and computational overhead need to be comprehensively considered to achieve a more optimal resource \linebreak  allocation~strategy.}

\subsubsection{Federated Learning-Based Resource Allocation Methods}
Federated learning, as a distributed machine learning paradigm, has demonstrated unique advantages in semantic communication for the IoV. By enabling distributed training and knowledge sharing, federated learning optimizes resource allocation and enhances system efficiency while preserving data privacy.

\hl{Ref.}~\cite{10416926} proposed a Mobility-Aware Split Federated Transfer Learning (MSFTL) framework, offering a novel approach to resource optimization in IoV semantic communication. The framework divides model training into four parts, leveraging split federated learning to reduce computational costs for vehicles. By incorporating a Stackelberg game-based resource optimization mechanism, it considers factors such as vehicle dwell time, computational load, and communication overhead to fairly select the most suitable training data volume for each vehicle and the entire network. This approach effectively reduces training costs and improves resource utilization efficiency. {Experiments show that MSFTL outperformed traditional Federated Learning (FL) in terms of convergence speed and final accuracy, with lower computational overhead. In the scenario of few-shot learning, it also has an advantage in terms of communication cost. However, its communication efficiency during large-scale data transmission still needs to be improved.}

\hl{Ref.}~\cite{10333738} introduced the Federated Vehicular Semantic Communication (FVSCom) framework, which also focuses on resource optimization. This framework employs federated learning for semantic extraction and proposes a semantic utility metric to evaluate performance. The problem of maximizing semantic utility is transformed into a stochastic optimization problem, which is solved using a deep reinforcement learning-driven dynamic semantic-aware algorithm. This approach achieves efficient semantic extraction and resource allocation, improving computational efficiency and semantic extraction accuracy while enhancing robustness to scenarios where vehicles leave or drop out. {However, the limitation of this study is that, although a semantic utility index is proposed, in the actual IoV scenarios, key performance indicators such as the specific transmission delay and computational overhead under different channel conditions are not taken into account.}

\textcolor{black}{Through comparative analysis, it can be found that although \cite{10416926,10333738} both focus on the resource optimization problem in semantic communication of the IoV, they have different emphases.The MSFTL framework, through split federated learning and a resource optimization mechanism based on the Stackelberg game, performs excellently in reducing the computational cost of vehicles and improving the efficiency of resource utilization. Especially in the scenario of few-shot learning, it has an obvious advantage in communication cost. However, its communication efficiency is insufficient during large-scale data transmission.The FVSCom framework, by introducing a semantic utility index and a dynamic semantic-aware algorithm driven by deep reinforcement learning, achieves efficient semantic extraction and resource allocation, improves computational efficiency and the accuracy of semantic extraction, and also enhances the robustness in the situation of vehicles leaving or exiting. However, it does not adequately consider key performance indicators such as transmission delay and computational overhead under different channel conditions.These two methods each have their own advantages in terms of resource allocation optimization, but they also have certain limitations. Future research can consider combining the advantages of the two while making up for their respective deficiencies so as to further improve the efficiency and performance of resource allocation in semantic communication of the IoV.
}


\subsection{\textcolor{black}{Data Security and Privacy Protection}}

\textcolor{black}{The application of semantic communication in the IoV has significantly improved communication efficiency and intelligence levels. However, it has also introduced new security and privacy challenges. The sensitivity of semantic information (such as vehicle trajectories, driving intentions, passenger preferences, etc.) requires the system to ensure confidentiality, integrity, and reliability during the transmission process. This section systematically analyzes the security risks faced by semantic communication in the IoV and summarizes the existing countermeasures.}

\subsubsection{\textcolor{black}{Security Risk Analysis}}

\textcolor{black}{The security threats to semantic communication systems mainly manifest as adversarial attacks, privacy leakage, man-in-the-middle attacks, model poisoning attacks, etc. Their risk characteristics and attack mechanisms show significant differences:}
{\begin{enumerate}
		\item	Adversarial attacks: Existing research shows that semantic communication systems are significantly vulnerable to adversarial attacks. Tiny perturbations generated based on algorithms such as Auto-PGD, FSGM, and DeepFool can lead to a substantial decline in the accuracy of semantic segmentation~\cite{10595916}. Of particular concern is the new type of covert attack mechanism. For example, the Covert Semantic Backdoor Attack (CSBA) can achieve the directional elimination of target semantics (such as traffic signs) without explicit triggers by analyzing the self-contained semantic features of the transmitted images~\cite{10598360}. Experiments show that even under high SNRs, the CSBA can still successfully remove the target semantics, and the restored image is visually indistinguishable from the original image. In addition, the Semantic Noise Attack (SNA) can inject semantic-level interference into the transmitted data, causing cascading error propagation in the encoding and decoding stages and leading to the failure of system decision making~\cite{10251891}. 
		\item	Privacy leakage risk: The deep correlation characteristics of semantic information enable attackers to reverse-derive users' sensitive data through multidimensional semantic analysis. For example, by analyzing the spatio-temporal patterns of vehicle trajectory semantics, users' resident areas and travel patterns can be inferred. Continuous monitoring of driving intention semantics may expose confidential information such as commercial transportation routes.
		\item	Man-in-the-middle attack threat: In V2V/V2I communication links, attackers can take advantage of the vulnerabilities of semantic protocols to conduct data eavesdropping and tampering. Typical attack scenarios include forging emergency braking commands, tampering with the semantic state of traffic lights, and hijacking path planning semantic data to induce vehicles to enter a preset area. Since such attacks directly operate on semantic layer information, traditional encryption mechanisms are difficult to effectively detect them.
		\item	Model poisoning attack: During the construction of a distributed semantic knowledge base, malicious vehicles can carry out covert poisoning by uploading contaminated data (such as distorted semantic features of traffic signs). More seriously, the poisoning attack may trigger systematic deviations in the semantic rule system, resulting in the failure of the Vehicle-to-Everything (V2X) collaborative decision-making mechanism.
\end{enumerate}}

\textcolor{black}{The current defense mechanisms mainly focus on single-point protection, and a security system covering the entire chain of semantic generation, transmission, and processing has not been formed yet. In particular, there are still significant technical gaps in detection algorithms for semantic-level covert attacks and dynamic privacy protection solutions, which urgently require systematic breakthroughs through the design of cross-layer defense architectures and lightweight cryptographic primitives.}

\subsubsection{\textcolor{black}{Countermeasures}}

\textcolor{black}{To effectively address the numerous security risks faced by semantic communication in the IoV, researchers have proposed a series of comprehensive countermeasures and key technologies. These can be addressed through methods such as semantic information encryption, federated learning, blockchain, edge intelligence, and adversarial \linebreak  sample~detection, which are defined below:}
{\begin{enumerate}
		\item	Semantic information encryption: Semantic information encryption is an important means to protect the confidentiality of semantic data in the Internet of Vehicles. By designing lightweight semantic-aware encryption algorithms, such as semantic feature obfuscation technology based on lattice cryptography, end-to-end confidentiality can be achieved while ensuring semantic decodability. This encryption method can effectively prevent data from being illegally stolen and tampered with during data transmission and storage, ensuring the security of semantic information. 
		\item	Federated learning and differential privacy: Federated learning is a distributed machine learning framework that enables distributed training of semantic models without sharing the original data. Combined with differential privacy technology, by adding controllable noise to semantic features, the leakage of original data can be further prevented. This combined approach can not only protect data privacy but also improve the robustness and generalization ability of semantic models. For example, \hl{refs.}~\cite{10416926,10333738} have elaborated on the application of federated learning in semantic communication of the Internet of Vehicles, demonstrating its remarkable effects in privacy protection and model performance improvement.
		\item	Blockchain and edge intelligence: The introduction of blockchain technology provides new ideas for data security and privacy protection in semantic communication of the Internet of Vehicles. Using blockchain to record the update operations of the semantic knowledge base can ensure the consistency and immutability of semantic rules. For example, the blockchain sharding technology proposed in \cite{LIN2024} reduces the verification delay of the knowledge base by dividing the knowledge base into multiple small pieces for verification while effectively resisting tampering attacks. In addition, the application of edge intelligence also provides strong support for privacy protection. \hl{Ref.}~\cite{feng2024semantic} achieved a balance between semantic understanding and vehicle privacy by building shared and private knowledge bases on edge servers. The shared library stores the background knowledge of autonomous driving on the edge server, while the private library stores the unique information of vehicles. The private library can be transmitted to the edge server according to the travel plan and updated by the vehicle itself when updated, and the edge server aggregates multisource information to update the shared library. This hierarchical architecture allows vehicles to only update and maintain their own private knowledge bases without uploading all data to the shared knowledge base, thus greatly reducing the risk of data leakage.
		\item	Adversarial sample detection: Adversarial sample detection is an important technology to deal with potential attacks in semantic communication of the Internet of Vehicles. The semantic anomaly detection module constructed based on the generative adversarial network (GAN) can identify adversarial semantic features in real time. \hl{Ref.}~\cite{10251891} proposed a defense mechanism based on Semantic Distance Minimization (SDM). SDM generates adversarial samples during the training process and optimizes the model to enable it to extract correct semantic information from adversarial samples. This method not only improves the model's robustness against adversarial attacks but also enhances the model's semantic understanding ability to a certain extent, ensuring the accuracy and reliability of semantic communication.
	\end{enumerate}
}

\textcolor{black}{In conclusion, through the comprehensive application of key technologies such as semantic information encryption, federated learning and differential privacy, blockchain and edge intelligence, and adversarial sample detection, the issues of data security and privacy protection in the combination of the Internet of Vehicles and semantic communication can be effectively addressed. These methods are not only innovative in theory but also show good effects in practical applications, providing a solid technical guarantee for the further development of semantic communication in the Internet of Vehicles.}

\subsection{\textcolor{black}{Lessons Learned}}

\textcolor{black}{Semantic information extraction and communication architectures must balance computational efficiency with accuracy. Deep learning-based methods (e.g., Swin Transformer, GANs, etc.) excel in multimodal fusion and noise resistance but face deployment challenges due to high resource demands. Reinforcement learning and federated learning show promise in dynamic resource allocation but require optimization for real-time IoV constraints. Blockchain integration enhances security but introduces scalability issues. Future work should prioritize lightweight models, edge-compatible algorithms, and hybrid optimization strategies to address these trade-offs.}

\section{Applications of Semantic Communication in IoV}\label{sec4}
\textcolor{black}{In Section \ref{sec3},
~we delved deep into the key technologies of semantic communication in the IoV, including semantic information extraction, communication architecture design, and resource allocation and management. These technologies form the foundation for achieving efficient and reliable semantic communication. They involve how to extract crucial semantic information from raw data, how to design the communication architecture to support the transmission of semantic information, and how to manage and allocate resources to optimize communication performance.The design of the semantic communication architecture needs to take into account the requirements of different application scenarios. For example, intelligent driving decision support may require communication with low latency and high reliability, while traffic management may place more emphasis on the comprehensiveness and accuracy of data. In this chapter, we will demonstrate how these key technologies function in practical application scenarios and how they support key applications in the IoV, such as intelligent traffic management, driving decision support, and service optimization. Through these application cases, we will further verify the potential of semantic communication in the IoV and explore its advantages and challenges in actual deployment.}
Table ~\ref{tab5} summarizes several application scenarios of semantic communication in the IoV.

\begin{table}[H]
\small
	\caption{\hl{Application} 
 scenarios of semantic communication in IoV.\label{tab5}}
		\begin{adjustwidth}{-\extralength}{0cm}
				\newcolumntype{C}{>{\centering\arraybackslash}X}
				\begin{tabularx}{\fulllength}{cCC}
				\toprule
				\textbf{Application Scenarios}	& \textbf{Representative Research }	     & \textbf{Advantages}\\
				\midrule
				\multirow[m]{5}{*}{Traffic Environment Perception}	& Environmental Semantic Communication~\cite{imran2024environment}				& Reduces data transmission volume, improves system response ability, and is suitable for millimeter-wave and terahertz communication systems.\\
				\cmidrule{2-3}
				& Cooperative Perception Semantic Communication~\cite{sheng2024semantic}					& Improves perception accuracy and throughput in low-signal-to-noise-ratio environments and avoids the ``cliff effect''.\\

				\bottomrule
\end{tabularx}
\end{adjustwidth}
\end{table}

\begin{table}[H]\ContinuedFloat
\small
\caption{{\em Cont.}}
\begin{adjustwidth}{-\extralength}{0cm}
				\newcolumntype{C}{>{\centering\arraybackslash}X}
				\begin{tabularx}{\fulllength}{cCC}
				\toprule
				\textbf{Application Scenarios}	& \textbf{Representative Research }	     & \textbf{Advantages}\\
				\midrule

			\multirow[m]{4.5}{*}{Intelligent Driving Decision Support}    & Dynamic Roadblock Semantic Traffic Control~\cite{10622833}						& Improves decision-making accuracy and real-time performance, and reduces communication overhead.\\
				\cmidrule{2-3}
				& High-altitude Platform Semantic Communication~\cite{10049005}					& Reduces communication costs; improves overall system performance and decision-making accuracy.\\
			\midrule		
				
				\multirow[m]{6.5}{*}{IoV Service Optimization}    & SemCom-empowered Service Provisioning Scheme~\cite{10271127}					& Significantly reduces queuing delay and improves the throughput of semantic data packets.\\
				\cmidrule{2-3}
				& Receiver-demand-centered Semantic Communication System~\cite{liu2024receivercentricgenerativesemanticcommunications}					& Greatly reduces data transmission volume and improves users' personalized experience.\\
				\cmidrule{2-3}
				& 6G Semantic Communication Scheme~\cite{9788561}					& Improves communication efficiency and service quality in in-vehicle scenarios.\\
				\midrule
				\multirow[m]{13}{*}{Intelligent Traffic Management}    & Vehicle Quantity Prediction Model~\cite{10571211}					& Improves the decision-making support ability for traffic signal control and congestion alleviation.\\
				\cmidrule{2-3}
				& Scalable Multitask Semantic Communication System (SMSC-FIR)~\cite{10095672}					& Improves adaptability under dynamic channel conditions and significantly improves multitask processing efficiency.\\
				\cmidrule{2-3}
				& Diffusion Model-based Channel Enhancer (DMCE)~\cite{zeng2024dmce}					& Improves the channel interference suppression ability of multiuser semantic communication systems and enhances the quality of semantically segmented images.\\
				\cmidrule{2-3}
				& Emergency Vehicle Dispatching Semantic Communication~\cite{10719121}					& Improves the passage efficiency of emergency vehicles and reduces interference with other traffic flows.\\
				\bottomrule
			\end{tabularx}
			\end{adjustwidth}
	\end{table}

\subsection{Traffic Environment Perception and Understanding}
In the IoV, vehicles need to perceive and understand the surrounding traffic environment in real time, including road conditions, traffic signs, and the dynamics of other vehicles and pedestrians. This is crucial for enhancing traffic safety, optimizing traffic flow, and enabling advanced applications such as autonomous driving. Traditional data transmission methods struggle to meet the demands of massive, real-time, and dynamically changing traffic information. Semantic communication, with its deep mining and precise transmission of data semantics, offers a novel solution for traffic environment perception and understanding in IoV. The following research cases demonstrate the effectiveness of semantic communication in this field.

The deployment of distributed sensing nodes in the IoV can significantly enhance the coverage and accuracy of environmental perception. \hl{Ref.}~\cite{imran2024environment} proposed an environmental semantic communication method to support distributed sensing-assisted networks, particularly in millimeter-wave (mmWave) and terahertz (THz) communication systems. By deploying multiple distributed sensing nodes, the study extracts environmental semantic information (e.g., bounding boxes and masks of targets) using RGB cameras and transmits this information to base stations to predict optimal beams. This approach significantly reduces the storage and transmission requirements of raw image data while improving the system's adaptability to dynamic environments. \textcolor{black}{The experiment was conducted based on the DeepSense 6G dataset, and the results show that the proposed solution can accurately predict the optimal beam in the real communication environment while reducing the transmission overhead of sensing data.}

In autonomous driving scenarios, collaborative perception among vehicles can provide more comprehensive environmental information, thereby enhancing driving safety. \hl{Ref.}~\cite{sheng2024semantic} proposed a semantic communication framework that combines Joint Source--Channel Coding (JSCC) and Hybrid Automatic Repeat Request (HARQ) technologies. The framework extracts critical semantic features through importance maps and optimizes transmission. By employing intermediate fusion, sensor data (e.g., LiDAR point clouds) from vehicles are semantically processed and transmitted using Orthogonal Frequency Division Multiplexing (OFDM) in time-varying multipath fading channels. \textcolor{black}{The OPV2V dataset was used in the study. Simulation results show that, compared to the traditional separate source--channel coding method, the proposed model has significantly improved in terms of sensing performance and throughput. In addition, this study also introduced a new semantic error detection method, SimCRC, and combined it with Hybrid Automatic Repeat Request (HARQ) to enhance the transmission reliability in scenarios with low SNRs. }

\textcolor{black}{Both of the above two experimental studies verify the effectiveness of the proposed methods through specific experimental designs. The work of \hl{Ref.}~\cite{imran2024environment} performed outstandingly in reducing data transmission overhead and beam prediction, and it is more suitable for optimizing communication resources. However, as the amount of data increases, this requires a higher processing capacity of the base station. The work of \hl{Ref.}~\cite{sheng2024semantic} has obvious advantages in sensing performance, and indicators such as AP@0.5 and AP@0.7 are superior to traditional methods. It is also more suitable for scenarios of precise target recognition. However, as the amount of data increases, it may face the problem of insufficient computing resources.}

\textcolor{black}{Through the above discussion on traffic environment perception and understanding, we can see the importance of the semantic information extraction techniques proposed in Chapter 3 in practical applications. These techniques not only improve the accuracy of environmental perception but also optimize communication efficiency by reducing the amount of data transmitted. In addition, the design of the semantic communication architecture ensures the reliable transmission of information in dynamic and complex traffic environments. The resource allocation and management strategies further optimize the transmission of semantic information, ensuring that critical information can be preferentially processed and transmitted under limited bandwidth and computing resources. The integrated application of these techniques demonstrates the potential of semantic communication in enhancing the performance of the IoV.}

\textcolor{black}{In the scenarios of achieving traffic environment perception and understanding, vehicles must be capable of processing and integrating information from different sensors and data sources in real time. The challenge in this process lies in the fact that the data formats and accuracies of different sensors vary, and the traffic environment itself is dynamically changing, including constantly changing weather conditions and traffic flows. These factors not only increase the complexity of data fusion but also require the perception system to have a high degree of adaptability and accuracy to ensure that it can still provide reliable environmental understanding in a changeable environment.}

\subsection{Intelligent Driving Decision Support}
In intelligent driving scenarios, vehicles need to perceive the surrounding environment in real time and make rational decisions, which relies on accurate and efficient information transmission and processing. Semantic communication technology can precisely extract key semantic information from massive traffic data, avoiding the transmission of redundant data in traditional communication methods, thereby significantly improving data transmission efficiency and decision-making speed. Below are several research cases that utilize semantic communication to enhance intelligent driving decisions.

In autonomous driving scenarios, vehicles need to perceive dynamic changes in the environment, such as sudden road obstacles (e.g., road maintenance, accidents, or vehicle breakdowns). In such cases, vehicles must quickly decide whether to change lanes or maintain their current lane to ensure safe and efficient driving. \hl{Ref.}~\cite{10622833} proposed a deep learning-based semantic traffic control system that assigns semantic encoding tasks to vehicles themselves rather than relying on server processing, thereby alleviating resource constraints. Specifically, autonomous vehicles (AVs) collect critical driving dynamics (e.g., speed, acceleration, and position) through sensors and use the deep Q-network (DQN) algorithm for decision making. The system converts complex driving environment information into compact semantic representations through semantic encoding and transmits them to the traffic monitoring module. The traffic monitoring module receives the encoded information and processes it using the DQN algorithm to derive appropriate driving decisions. \textcolor{black}{In the experiment, the researchers used the KUL traffic sign classification dataset from Belgium, the MASTIF dataset from Croatia, and the German Traffic Sign Recognition Benchmark dataset to train the object detection model through a custom framework. The experimental results show that, compared to directly transmitting images, when the semantic system transmits a single image, 900 images equivalent to a 30 s video, and 1800~images equivalent to a 60 s video, the file transmission time is significantly shortened, and the file size is also greatly reduced. In addition, the accuracy rate of the traffic optimization model of this system reached 92\%. Although affected by the noise added in the simulation (such as changes in vehicle speed), this result still indicates that the system has a high degree of reliability in intelligent driving decision making.}


In more complex traffic scenarios, such as fully connected autonomous vehicle networks involving High-Altitude Platforms (HAPs), semantic communication also plays a crucial role. \hl{Ref.}~\cite{10049005} proposed an AI-driven semantic communication framework for this scenario. In this framework, the traffic infrastructure (TI) extracts semantic concepts from traffic signs using a convolutional autoencoder (CAE) and transmits these concepts to macro base stations (MBSs). Upon receiving the semantic concepts, the MBS uses the Proximal Policy Optimization (PPO) algorithm to make decisions for CAVs. For example, when the TI observes a ``left turn ahead'' sign, it extracts the corresponding semantic concept and transmits it to the MBS. The MBS analyzes the concept using the PPO algorithm and generates appropriate driving instructions, such as ``turn left'', for the CAVs. \textcolor{black}{In the experiment, 12 traffic signs from the TSRD dataset were used, and the proposed framework was compared with the Augmented Random Search (ARS) baseline. The results show that the deep Q-network (DQN) outperform ARS in terms of training convergence and received rewards, with the total reward increased by 37.11\%. In addition, as the size of the Resource Block (RB) increased, the total reward received by the Connected and Automated Vehicle (CAV) also increased, which indicates that more complete semantic information helps to improve the accuracy of decision making. This framework was able to reduce the communication cost by up to 90.81\%, demonstrating great potential in improving communication~efficiency.}

\textcolor{black}{Overall, both semantic communication methods have achieved positive results in the experiments, but some key factors still need to be considered in actual deployment. The method in \cite{10622833} shows high real-time performance and decision-making accuracy when dealing with sudden roadblock scenarios, making it suitable for autonomous driving scenarios with high requirements for response speed. However, its stability may be affected by the quality of model training data and the network environment. The framework \mbox{in \cite{10049005}} has significant advantages in reducing communication costs and improving resource utilization, and it is suitable for large-scale data transmission and decision support in the Internet of Vehicles. But in practical applications, it is necessary to address the resource requirements for deep Q-network (DQN) training and improve the system's adaptability to complex traffic environments.}

\textcolor{black}{In the context of exploring decision support for intelligent driving, the semantic communication architecture discussed in Section \ref{sec3} demonstrates its core role in handling real-time decision-making information. Semantic information extraction technology can filter out key information from the massive data generated by vehicles and their environments, such as the positions and speed changes of potential obstacles, which are crucial for the immediate response of vehicles. In addition, through optimized resource allocation and management strategies, it is ensured that these key pieces of information can be transmitted and processed in the shortest possible time, thereby supporting vehicles in making quick and accurate driving decisions. The combination of this technology and application not only improves the reliability of intelligent driving systems but also provides strong technical support for the development of future autonomous driving technologies.}

\textcolor{black}{In the scenario of intelligent driving decision support, vehicles need to quickly and accurately extract key semantic information from complex traffic data and make timely driving decisions based on it. The challenge in this process is how to ensure that in a dynamic and unpredictable traffic environment the decision-making system can not only accurately understand environmental information but also respond quickly to avoid potential collisions and improve driving safety. In addition, how to effectively utilize vehicle data while protecting personal privacy is also an issue that needs to be addressed.}

\subsection{IoV Service Optimization}
In the field of the IoV, leveraging semantic communication technology to enhance service performance has become a critical research direction. Semantic communication, with its advantages of reducing data volume and improving real-time performance and accuracy, offers users more efficient and personalized service experiences. Below, specific research cases are analyzed to delve into its applications in IoV service optimization.

The demand for next-generation ultra-reliable low-latency communication (xURLLC) in the IoV is growing, necessitating the optimization of semantic data packet transmission to reduce queuing delays and enhance user experience. \hl{Ref.}~\cite{10271127} addressed the challenges posed by resource scarcity and xURLLC requirements by proposing a SemCom-empowered Service Supplying Solution. Based on queuing theory, the study derives a formula for queuing delays of semantic data packets based on knowledge base matching. An optimization problem is formulated to minimize queuing delays, which is subject to reliability constraints in knowledge base construction (KBC) and vehicle service pairing (VSP). Using the Lagrangian dual method, the original problem is decomposed into two subproblems: in the first stage, a heuristic algorithm is applied to solve the KBC subproblem for potential vehicle pairs while identifying the optimal knowledge base construction strategy; in the second stage, based on the results of the first stage, a greedy algorithm is used to determine the optimal pairing strategy for the VSP subproblem. This solution significantly reduces queuing delays, improves the throughput of semantic data packets, and enhances user satisfaction. While the theoretical advantages of this approach are evident through precise mathematical derivations and algorithm design, its practical application may incur high computational costs and require advanced hardware performance, limiting its use in low-configuration vehicles.

Traditional semantic communication often fails to meet users' personalized needs due to the loss of critical receiver information. \hl{Ref.}~\cite{liu2024receivercentricgenerativesemanticcommunications} proposed a receiver-centric semantic communication system where the received component 
can request specific semantic information from senders. The sender utilizes the natural language processing capabilities of large language models (e.g., GPT-4) to understand the request, combined with tools such as object detection and license plate detection, to analyze video or image data. Through an attention mechanism-based semantic extraction algorithm, relevant semantic information is accurately extracted and fed back in text form. If task planning is unreasonable, a task reflection module replans or selects the most relevant video frames for transmission. This system emphasizes user-centric personalized demands, leveraging advanced language models and intelligent algorithms to address the shortcomings of traditional semantic communication. However, reliance on large language models introduces challenges, such as high training costs and potential delays in real-time scenarios due to model processing speed limitations, affecting service immediacy.

In in-vehicle scenarios, efficient perception and interaction are essential to enhance user experience. \hl{Ref.}~\cite{9788561} ~proposed a 6G semantic communication solution integrated with smart fabrics. Smart fabrics perceive the position, movements, and environmental conditions (e.g., temperature and humidity) of individuals inside the vehicle through principles such as electromagnetic induction, collecting raw data. The semantic sensing terminal designs a deep network based on recurrent neural networks (RNNs) and their variants (e.g., LSTM, GRU, etc.) to extract environmental data from time-series features of the data stream, thus reducing transmission data volume through dimensionality reduction techniques. Semantic information is transmitted via an integrated space--air--ground network to a remote AI platform, where technologies like GANs enable semantic recovery and advanced deep learning services. This solution achieves high-quality signal reconstruction and advanced services under different SNR conditions, effectively improving communication efficiency and service quality in in-vehicle scenarios. While this solution is specifically designed for in-vehicle scenarios, where 6G technologies and smart fabrics are leveraged to provide high-quality communication and service experiences, its applicability is limited to other IoV scenarios, such as vehicle-to-vehicle or vehicle-to-infrastructure communication.

\textcolor{black}{By comparing the experimental results of the three literature sources, it can be seen that the S4 scheme in \cite{10271127} performs excellently in the service optimization of the Internet of Vehicles. It has significant advantages, especially in reducing latency and increasing throughput, and is suitable for Internet of Vehicles scenarios with high real-time requirements. The system in \cite{liu2024receivercentricgenerativesemanticcommunications} performs well in reducing the amount of data transmission and meeting the semantic information needs of the receiving end, and it is suitable for scenarios that require efficient transmission of key information. However, there is still room for improvement in its performance when dealing with complex requests. The DL-SCMT model in \cite{9788561} performs outstandingly in terms of signal reconstruction and classification accuracy and has good robustness, especially in environments with a low signal-to-noise ratio. It is suitable for 6G application scenarios that require high reliability and intelligent interaction.}

\textcolor{black}{The discussion on the service optimization of the IoV further confirms the potential of semantic communication in enhancing the service quality and user experience. Through identifying the true intentions behind user requests, semantic information extraction technology enables service providers to more accurately meet user needs. The optimized strategies for resource allocation and management ensure the efficient transmission of semantic information between users and service providers, reducing unnecessary data transmission and improving the service response speed. This user-centered service optimization approach not only enhances user satisfaction but also provides a new direction for the innovation and development of IoV services.}

\textcolor{black}{In the scenario of IoV service optimization, the challenge lies in how to meet users' demands for personalized services while maintaining the efficiency and consistency of services. As users' expectations for IoV services continue to rise, service providers need to optimize resource allocation under limited network resources to support services with different quality of service requirements. In addition, integrating diverse services into a unified platform while ensuring service interoperability and an excellent user experience is also a technical challenge.}


\subsection{Intelligent Traffic Management}
In the field of intelligent traffic management, semantic communication has demonstrated significant potential and value. Numerous research papers have focused on this area, providing rich insights and practical outcomes for the application of semantic communication. Below, we explore the specific applications of semantic communication in intelligent traffic management, including vehicle count prediction, multitask communication, and emergency vehicle scheduling.

Vehicle count prediction is a critical component of intelligent traffic management to provide decision-making support for traffic signal control and congestion mitigation. \hl{Ref.}~\cite{10571211} proposed a vehicle count prediction model based on semantic communication. This model employs a joint convolutional neural network (CNN) and long short-term memory network (LSTM) to construct a semantic encoder--decoder architecture. Raw images captured by cameras are first processed by the semantic encoder to extract key semantic information, such as vehicle density maps, which are then converted into symbols and transmitted to the central traffic controller (CTC). The CTC utilizes the received semantic information, combined with the LSTM's ability to handle temporal correlations in image sequences, to achieve accurate vehicle count predictions for various road segments. Based on these predictions, scientifically effective traffic management strategies are formulated. \textcolor{black}{The experiment utilized the TRANCOS dataset, which consists of 1,244 images, and a total of 46,796 cars were labeled. The results show that, compared to the traditional source encoder/decoder method, this model reduced the communication overhead by 54.42\%. In terms of the Mean Absolute Error (MAE) and Mean Squared Error (MSE), this model outperformed the existing methods based on GRU, LSTM, and FCN-rLSTM respectively. The MAE was reduced by 90.71\%, 73.03\%, and 19.1\%, respectively, and the MSE was reduced by 103.91\%, 77.74\%, and 13.45\%, respectively. This result indicates that semantic communication technology can significantly improve the accuracy of vehicle number prediction while reducing the amount of data transmission, providing strong support for real-time traffic~management.}

As the complexity of intelligent transportation systems increases, multitask processing capabilities have become increasingly important. \hl{Ref.}~\cite{10095672} discussed a scalable multitask semantic communication system (SMSC-FIR) that focuses on tasks such as vehicle reidentification (ReID), vehicle color classification, and vehicle type classification. The system prioritizes the transmission of critical semantic features using a Feature Importance Ranking (FIR) method. \textcolor{black}{The experiment made use of the VeRi-776 dataset, which contains more than 50,000 images of 776 vehicles. The results demonstrate that under low-SNR conditions, the SMSC-FIR outperformed the existing state-of-the-art methods in various tasks. For instance, at 0 dB, compared to the sequential selection and random selection methods, the performance of vehicle reidentification (ReID) was improved by 40.0\% and 212.2\%, respectively, the performance of color classification was enhanced by 11.0\% and 9.8\%, respectively, and the performance of type classification was increased by 6.9\% and 7.0\%, respectively. This result indicates that through feature importance ranking and dynamic coding rate adjustment, SMSC-FIR can effectively improve task performance in complex multitask scenarios, especially when the channel conditions are poor.}

In multiuser semantic communication systems, channel interference remains a critical issue affecting communication quality. \hl{Ref.}~\cite{zeng2024dmce} proposed a Diffusion Model-based Channel Enhancer (DMCE) to address channel interference in multiuser semantic communication systems for intelligent traffic management. The DMCE improves channel equalization performance by learning the specific data distribution of channel effects on transmitted semantic features, thereby suppressing noise in channel state information (CSI) estimation. In the system model, multiple users transmit semantic features of multisource images (e.g., RGB and infrared images) of the same traffic scene through a MIMO channel to a centralized receiver. The receiver then recovers and fuses the semantic features using channel equalization and DMCE-enhanced CSI estimation while ultimately generating a semantic segmentation image of the traffic scene. \textcolor{black}{The experiment used a multisource image dataset containing 1,569 pairs of RGB-IR urban traffic scene images. The results show that under low-SNR conditions, the DMCE could increase the mean Intersection over Union (mIoU) by more than 25\%. In addition, the DMCE significantly reduced the Normalized Mean Squared Error (NMSE) of the channel state information (CSI) estimation, with an average reduction of 14 dB. This result indicates that the DMCE can effectively improve the accuracy of semantic feature recovery in complex wireless channel environments, providing strong support for the practical application of multiuser semantic communication systems.}

Efficient scheduling of emergency vehicles is crucial for improving emergency response capabilities. \hl{Ref.}~\cite{10719121} explored the application of semantic communication in 6G networks, particularly for resource optimization in emergency vehicle scheduling. Through semantic encoding and decoding, data from emergency vehicles (e.g., GPS location, speed, and medical equipment information) are converted into semantic information and sent to a traffic infrastructure that dynamically adjusts traffic signal durations to prioritize emergency vehicles. This semantic communication-based scheduling method not only improves the efficiency of emergency vehicles but also minimizes disruption to other traffic flows, thus significantly enhancing emergency response capabilities. However, the scalability of this solution in large-scale urban traffic networks poses challenges, because how to ensure efficient operation and rapid response in increasingly complex and high-traffic environments remains a key focus for future research.

\textcolor{black}{In these four studies, \hl{ref.}~\cite{10571211} verified the advantages of the semantic communication model based on CNN-LSTM in reducing overhead and improving prediction accuracy by constructing the model and conducting training and testing on the TRANCOS dataset. \hl{Ref.}~\cite{10095672} proposed the SMSC-FIR system and carried out multitask learning experiments on the VeRi-776 dataset. Through comparison with existing methods, it demonstrated the performance improvement of the system under low-SNR conditions. \hl{Ref.}~\cite{zeng2024dmce} proposed the DMCE scheme. By introducing a diffusion model into the multiuser semantic communication system to enhance the estimation of channel state information, it improves the accuracy of semantic feature recovery. }

\textcolor{black}{In the application scenarios of intelligent traffic management, the key technologies of semantic communication introduced in Chapter 3 have demonstrated their significant value in improving the overall efficiency and safety of the traffic system. Semantic information extraction technology enables the traffic management system to quickly identify key events, such as traffic accidents or congestion situations, from complex traffic data. The semantic communication architecture supports the rapid and reliable transmission of this information, while the resource allocation and management strategies ensure the real-time nature and accuracy of traffic management decisions. The integrated application of these technologies not only enhances the intelligent level of traffic management but also provides technical support for the sustainable development of urban traffic.}

\textcolor{black}{In the scenario of intelligent traffic management, the system needs to handle large-scale data from an entire city or region, which poses challenges to data storage, processing, and analysis capabilities. Moreover, in order to achieve effective traffic control, the collaborative work among different traffic management systems is of vital importance. This requires the systems to be able to achieve effective coordination among different management platforms. At the same time, the implementation of intelligent traffic management may be restricted by existing policies and regulations. It is necessary to cooperate with the government and regulatory authorities to promote the necessary updates of regulations so as to adapt to the development and application of new technologies.}

\subsection{\textcolor{black}{Lessons Learned}}

\textcolor{black}{Semantic communication significantly improves IoV applications like traffic perception and emergency response by reducing data redundancy and enhancing decision accuracy. Case studies demonstrate that task-oriented semantic frameworks (e.g., HARQ-enhanced perception and generative AI-driven scheduling) outperform traditional methods in low-SNR scenarios. However, generalization to complex environments (e.g., extreme weather) remains a challenge. Collaborative systems leveraging edge computing and distributed knowledge sharing are critical for scalability, while privacy-preserving mechanisms (e.g., federated learning) must be prioritized in sensitive applications like remote diagnostics.}

\section{ Challenges and Future Research Directions}\label{sec5}

\subsection{ Knowledge Base Creation and Updating}
In the semantic communication-based IoV, the knowledge base plays an indispensable role as the foundation for achieving accurate semantic understanding and effective communication. However, the construction and updating of knowledge bases still face significant challenges. In terms of construction, the complexity of data sources is a primary obstacle. In the IoV, vehicles generate massive amounts of data through various onboard sensors, while external data sources such as traffic management centers and map service providers continuously contribute additional data. These data are often heterogeneous in format and lack unified standards, making integration highly challenging. Moreover, the inherent complexity and ambiguity of traffic-related knowledge pose difficulties in knowledge representation. For instance, describing traffic congestion is not a simple binary state of ``congested'' or ``clear''; it involves multiple degrees of congestion influenced by factors such as time, location, and traffic flow. Accurately representing such complex knowledge in a machine-understandable form remains a major challenge. Selecting appropriate knowledge representation methods, such as semantic networks, ontologies, or production rules, is crucial for transforming complex traffic knowledge into a usable format. In semantic communication, the knowledge base must be dynamically updated based on data from sources, channels, and tasks. The dynamic nature of IoV also imposes stringent real-time requirements on knowledge base updates. However, current update mechanisms often suffer from delays and cannot meet the real-time demands of the IoV.

To address these challenges, future research can explore multidimensional approaches by integrating cutting-edge technologies. On the one hand, incorporating AI large models can leverage their powerful natural language processing and knowledge understanding capabilities to deeply analyze and integrate heterogeneous traffic data, efficiently constructing knowledge graphs. The continuous learning capabilities of large models can also facilitate real-time updates to the knowledge base, enabling them to promptly capture changes in the traffic environment. On the other hand, federated learning can be employed to enable collaborative training of distributed data while preserving data privacy, thereby enriching the knowledge base and enhancing its generalizability. Additionally, blockchain technology can be introduced to ensure the security and consistency of the updating process. By leveraging its tamper-proof and decentralized characteristics, blockchain can record update operations, ensuring reliable knowledge transmission. These combined efforts can drive the advancement of knowledge base technologies in the IoV.\textcolor{black}{In addition, the uncertainty of user behavior brings additional challenges to the update of the knowledge base. Users' personalized needs and behavior patterns change over time, which requires the system to be able to continuously learn and adapt to these changes to ensure that the relevant information in the knowledge base can reflect users' latest preferences and demands.}

\subsection{Semantic Understanding and Ambiguity}
The IoV involves multisource heterogeneous data from various vehicles, sensors, and infrastructure, which differ in format, semantics, and precision. For example, sensors from different manufacturers may use varying measurement units and data formats, while data exchange between roadside infrastructure and vehicles often encounters semantic mismatches. Integrating and accurately understanding the semantics of such heterogeneous data is a significant challenge, requiring the development of universal data models and semantic transformation methods to achieve seamless integration and effective utilization. Additionally, natural language descriptions of traffic scenarios and instructions often contain ambiguities. For instance, the phrase ``slow down at the upcoming intersection'' lacks a clear definition of what constitutes ``slow'', leading to potential differences in interpretation among drivers or vehicle systems. In semantic communication, resolving such ambiguities is critical. This can be achieved by establishing precise semantic models, incorporating contextual information, and leveraging knowledge base reasoning to ensure accurate transmission and understanding of semantic information.\textcolor{black}{In the process of multisource data fusion, the heterogeneity of data and the dynamic changes of the environment may lead to ambiguities and misunderstandings of semantic information. Therefore, the development of advanced data fusion technologies capable of handling these complexities is the key to achieving accurate semantic understanding.}

To address these challenges, future research can focus on integrating advanced technologies to achieve seamless data integration and accurate semantic understanding. On the one hand, efforts should be directed toward developing universal data models and semantic transformation methods, including cross-domain unified models, dynamic semantic conversion, and international standardization. On the other hand, research should emphasize ambiguity resolution through context-aware semantic understanding, multimodal semantic fusion, and human--machine collaborative semantic interaction. Advanced technologies such as deep learning, reinforcement learning, and generative AI can be utilized for feature extraction, autonomous decision making, and data augmentation of heterogeneous data. Knowledge graph construction and fusion can provide background knowledge and reasoning support for semantic understanding. Additionally, edge computing can be employed to decentralize semantic processing, enabling intelligent decision making at the edge and enhancing the real-time performance and efficiency of semantic communication.

\subsection{Real-Time and Reliability Requirements}
In the IoV, vehicles are in constant high-speed motion, which leads to rapidly changing network topologies and unstable communication channels. This demands that semantic communication achieves real-time performance in highly dynamic environments to achieve timely transmission and processing of semantic information. For example, in scenarios such as emergency braking or lane changes, relevant semantic information (e.g., braking intent or lane change direction) must be quickly transmitted to surrounding vehicles to prevent collisions. Therefore, vehicles need to rapidly encode and decode large amounts of road condition and driving intent information during operation. When multiple vehicles communicate simultaneously, encoding and decoding delays may occur, affecting the timeliness and accuracy of information transmission. \textcolor{black}{The dynamically changing traffic environment poses higher requirements for the real-time performance of the system. The system must be able to quickly adapt to environmental changes, such as fluctuations in traffic flow and sudden road conditions, to ensure the continuity of communication and the timeliness of decision making.}

In order to meet the real-time and reliability requirements, it is necessary to optimize the communication protocol, adopt efficient channel access technology and resource allocation algorithms, and reduce the communication delay. For example, \hl{ref.}~\cite{wang2019novel} proposed the TDMA-based Capture-Aware MAC protocol (CT-MAC), which effectively improves broadcast reliability and channel utilization efficiency by optimizing frame length and utilizing capture effect. This shows that optimization combined with the MAC protocol can better cope with the challenges of high vehicle dynamics in the vehicle networking environment, thus improving real-time performance and reliability and providing a useful reference for the combination of vehicle networking and semantic communication. Meanwhile, semantic information in the IoV is critical for traffic safety, necessitating reliable transmission. Interference, fading, and noise in wireless channels can lead to data loss or errors. To enhance reliability, error control techniques (e.g., ARQ and FEC), multipath transmission, and redundancy mechanisms should be employed to ensure accurate and error-free delivery of semantic information under adverse channel conditions. Future research can focus on low-latency, high-accuracy semantic encoding and decoding algorithms, such as those based on the Transformer architecture, to improve information processing speed and meet the real-time and reliability requirements of the IoV.

\subsection{Security and Privacy Protection}
Semantic communication involves sensitive semantic information such as vehicle trajectories, driving behaviors, and passenger details, the leakage of which could pose significant threats to user privacy. For instance, attackers could analyze semantic communication data to infer a vehicle's travel routes and stop locations, thereby violating user privacy. If semantic information is intercepted or tampered, this could lead to erroneous vehicle decisions that may jeopardize driving safety. Additionally, the IoV encompasses numerous communication links, including vehicle-to-vehicle (V2V) and vehicle-to-infrastructure (V2I) connections. These links are vulnerable to security threats such as hacking, interference, and attacks like semantic message forgery, tampering, and man-in-the-middle attacks. Malicious actors could send false traffic condition information to mislead vehicles or alter semantic communication commands between vehicles and infrastructure, causing traffic disruptions or even accidents. While intelligent connected vehicles can alleviate spectrum pressure through semantic communication, deep learning-based semantic communication is susceptible to adversarial attacks~\cite{10595916}, backdoor attacks~\cite{10598360}, and other security threats. The IoV based on semantic communication also faces risks such as eavesdropping attacks, adversarial attacks, and poisoning attacks~\cite{10251891}.  

To address these challenges, it is essential to develop encryption algorithms specifically designed for semantic information, integrating digital signatures and identity authentication technologies to ensure the confidentiality, integrity, and authenticity of semantic communication. Dedicated security communication protocols for the IoV should be established, leveraging technologies such as blockchain and federated learning~\cite{10328182}, to build distributed trust mechanisms, preventing communication link attacks and data theft. \textcolor{black}{Existing encryption and detection algorithms, such as homomorphic encryption, often introduce high latency and are difficult to meet the millisecond response requirements of the IoV. It is necessary to design lightweight security protocols, such as dynamic key distribution mechanisms based on edge computing. And existing detection methods are mostly targeted at specific attack patterns (such as FGSM or PGD), with insufficient generalization ability for unknown attack types. Combining meta-learning with online adaptive mechanisms may be a potential solution.}
\textcolor{black}{The acceptance of new technologies by users also affects the effectiveness of privacy protection. By enhancing users' trust in the system, making the data processing procedures transparent and strengthening data security measures, we can boost users' confidence in privacy protection.}

\subsection{Standardization and Regulation}
Currently, different enterprises and institutions in the IoV industry may have different definitions and encoding methods for semantics, making seamless semantic communication between vehicles and infrastructure challenging. Moreover, the IoV involves multiple industries, including automotive, telecommunications, and transportation, each of which has its own semantic standards and business requirements, and this thus complicates the harmonization of semantic standards.  

Therefore, stakeholders such as automotive manufacturers, communication companies, and research institutions should collaborate to develop unified semantic standards for the IoV, clarifying semantic definitions, encoding rules, and data formats to ensure compatibility across different devices and systems. Active participation in international standardization efforts is also crucial to align domestic standards with global ones, fostering the coordinated development of the worldwide IoV industry.

\subsection{\textcolor{black}{Lessons Learned}}

\textcolor{black}{The semantic IoV ecosystem faces unresolved challenges in security (e.g., adversarial attacks), real-time knowledge base updates, and standardization. Dynamic environments demand adaptive semantic models with low-latency processing, while cross-domain semantic alignment requires unified ontologies. Future research should focus on AI-driven solutions (e.g., large language models for knowledge fusion), lightweight encryption for resource-constrained devices, and global standardization efforts to ensure interoperability. Collaborative frameworks integrating academia, industry, and policymakers will accelerate practical adoption.}

\section{Conclusions}\label{sec6}

This paper provides a comprehensive and in-depth review of semantic communication technologies in the IoV that covers key aspects ranging from foundational technical backgrounds to specific applications, challenges, and future directions. As the core of intelligent transportation systems, the IoV integrates multiple communication modes to achieve comprehensive connectivity between vehicles and their environment. However, traditional communication technologies face numerous bottlenecks in IoV applications. Semantic communication, with its unique advantages such as reducing redundant transmissions, enhancing communication efficiency and accuracy, and ensuring information security, offers innovative solutions to IoV communication challenges while demonstrating broad application prospects in the field. By enabling efficient extraction, transmission, and understanding of semantic information, semantic communication can significantly improve the communication efficiency, data processing capabilities, and intelligent decision-making support of the IoV, providing robust technical support for safer, more efficient, and smarter transportation systems.  

Nevertheless, the application of semantic communication in the IoV still faces many challenges. Future research needs to further explore these critical issues and propose more effective solutions by integrating advanced technologies such as artificial intelligence, edge computing, and blockchain. Additionally, strengthening interdisciplinary collaboration and fostering close ties between academia, industry, and application will accelerate the practical implementation of semantic communication technologies in the IoV and promote the comprehensive development of intelligent transportation systems.



\vspace{6pt} 





\authorcontributions{Conceptualization, S.Y. and Q.W.; methodology, S.Y.; investigation, S.Y.; data curation, S.Y.; writing---original draft preparation, S.Y. and Q.W.; writing---review and editing, P.F. and Q.F.; visualization, S.Y.; supervision, Q.W. All authors have read and agreed to the published version of the manuscript.}

\funding{This work was supported in part by the National Natural Science Foundation of China under Grant No. 61701197, in part by the National Key Research and Development Program of China under Grant No. 2021YFA1000500(4), and in part by the 111 project under Grant No. B23008.}

\institutionalreview{Not applicable.}

\informedconsent{Not applicable.}

\dataavailability{No new data were created or analyzed in this study. Data sharing is not applicable to this article.}

\conflictsofinterest{The author Qiang Fan was employed by the company Qualcomm. The remaining authors declare that this research was conducted in the absence of any commercial or financial relationships that could be construed as a potential conflicts of interest.} 





\begin{adjustwidth}{-\extralength}{0cm}

\reftitle{References}




\PublishersNote{}
\end{adjustwidth}
\end{document}